\documentclass[aps,prb,twocolumn,amsmath,superscriptaddress,amssymb]{revtex4}
\usepackage{graphicx}
\usepackage{enumitem}
\usepackage[colorlinks,bookmarks=false,citecolor=red,linkcolor=blue,urlcolor=blue]{hyperref}

\newcommand{\bea}{\begin{eqnarray}}          
\newcommand{\eea}{\end{eqnarray}}

\allowdisplaybreaks

\begin{document}

\title{Berry phase in the rigid rotor: the emergent physics of odd antiferromagnets }
\author{Subhankar Khatua}
\email{subhankark@imsc.res.in}
\affiliation{The Institute of Mathematical Sciences, HBNI, C I T Campus, Chennai 600113, India}
\author{R. Ganesh}
\email{r.ganesh@brocku.ca}
\affiliation{Department of Physics, Brock University, St. Catharines, Ontario L2S 3A1, Canada}
\date{\today}

\begin{abstract}
The rigid rotor is a classic problem in quantum mechanics, describing the dynamics of a rigid body with its centre of mass held fixed. It is well known to describe the rotational spectra of molecules. The configuration space of this problem is $SO(3)$, the space of all rotations in three dimensions. This is a topological space with two types of closed loops: trivial loops that can be adiabatically shrunk to a point and non-trivial loops that cannot. In the traditional formulation of the problem, stationary states are periodic over both types of closed loops. However, periodicity conditions may change if Berry phases are introduced. We argue that time-reversal-symmetry allows for only one new possibility -- a Berry phase of $\pi$ attached to all non-trivial loops. We derive the corresponding stationary states by exploiting the connection between $SO(3)$ and $SU(2)$ spaces. The solutions are anti-periodic over any non-trivial loop, i.e., stationary states reverse sign under a $2\pi$ rotation about any axis. Remarkably, this framework is realized in the low-energy physics of certain quantum magnets. The magnets must satisfy the following conditions: (a) the classical ground states are unpolarized, carrying no net magnetization, (b) the set of classical ground states is indexed by $SO(3)$, and (c) the product $N\times S$ is a half-integer, where $N$ is the number of spins and $S$ is the spin quantum number. We demonstrate this result in a family of Heisenberg antiferromagnets defined on polygons with an odd number of vertices. At each vertex, we have a spin-$S$ moment that is coupled to its nearest neighbours. In the classical limit, these magnets have coplanar ground states. Their quantum spectra, at low energies, correspond to `spherical top' and `symmetric top' rigid rotors. For integer values of $S$, we recover traditional rigid rotor spectra. With half-integer-$S$, we obtain rotor spectra with a Berry phase of $\pi$.
\end{abstract}
                     
\maketitle
\section{Introduction}
The rotation of a rigid body is a fundamental problem in classical and quantum mechanics. It is one of the early problems where quantum spectra could be worked out and compared against experiments. It laid the foundation for the field of microwave rotational spectroscopy
\cite{Bunker2006,Xu2011}, with applications ranging from laboratory organic chemistry to interstellar space\cite{Arunan2015}. The solutions of the quantum rigid rotor have been known since the work of Casimir in 1931\cite{casimir}. Over the decades, this problem has been expanded to include details such as asymmetries, centrifugal distortion, etc. In this article, we revisit this problem and show that it allows for a non-trivial Berry phase structure. This can be viewed as reframing the problem with a new boundary condition that is consistent with the underlying topology. We evaluate the resulting spectrum and suggest magnetic realizations where this Berry phase structure is realized. 

The Berry phase or geometric phase can play a strong role in quantum mechanics\cite{Cohen2019}. As a simple illustration, we consider a particle on a circle --  a one-dimensional space with periodic boundaries. Solving the Schr\"odinger equation leads to wave-like solutions. Periodic boundary conditions then restrict the particle to quantized levels. If the circle were threaded by a magnetic flux, the particle accrues an Aharonov-Bohm phase with every revolution. This changes the boundary condition and thereby, the spectrum of the particle. The symmetries of the system strongly constrain the Aharonov-Bohm phase. For example, time-reversal symmetry only allows for two values: $0$ or $\pi$. In the latter case, the wavefunction at a given point cannot be defined uniquely. It must necessarily be double-valued as it switches sign after one revolution. The Aharonov-Bohm phase is a particular example of a more general notion -- the Berry phase, which may arise even in the absence of external fields. For example, in crystalline solids, Bloch wavefunctions may accrue phases over loops in the Brillouin zone\cite{Hasan2010}. This phenomenon serves as the starting point for the field of topological insulators. These phases are strongly constrained by time-reversal symmetry, space group symmetries, etc. They may even give rise to wavefunctions that are multi-valued. In this article, we consider Berry phase structure in the rigid rotor, constrained by time-reversal symmetry. We discuss magnetic realizations where the Berry phase is intrinsic and can be switched on or off by changing the spin quantum number. Our results can be viewed in analogy with the `Haldane gap' in one-dimensional spin chains, where Berry phase effects lead to a qualitative change in the excitation spectrum\cite{Haldane1983,Affleck1989}. 
  
We present a class of magnetic systems as realizations of these ideas. We consider rotationally-symmetric magnets with Heisenberg couplings.   In the classical limit, they possess degenerate ground states that are related to one another by rotations. We focus on magnets where each classical ground state can be associated to a unique rotation operation. In such systems, the quantum problem shows a characteristic low-energy spectrum, precisely that of a rigid rotor. For certain system sizes and values of the spin quantum number $S$, this `emergent' rotor accrues a non-trivial Berry phase. This modifies the eigenvalues and degeneracies, resulting in a new spectrum that is markedly different from the traditional rigid rotor. We support these assertions with analytic arguments and exact diagonalization results on a family of antiferromagnets.

\section{The rigid rotor and the topology of $SO(3)$}
\label{sec.rrotor}
\begin{figure}
\includegraphics[width=3in]{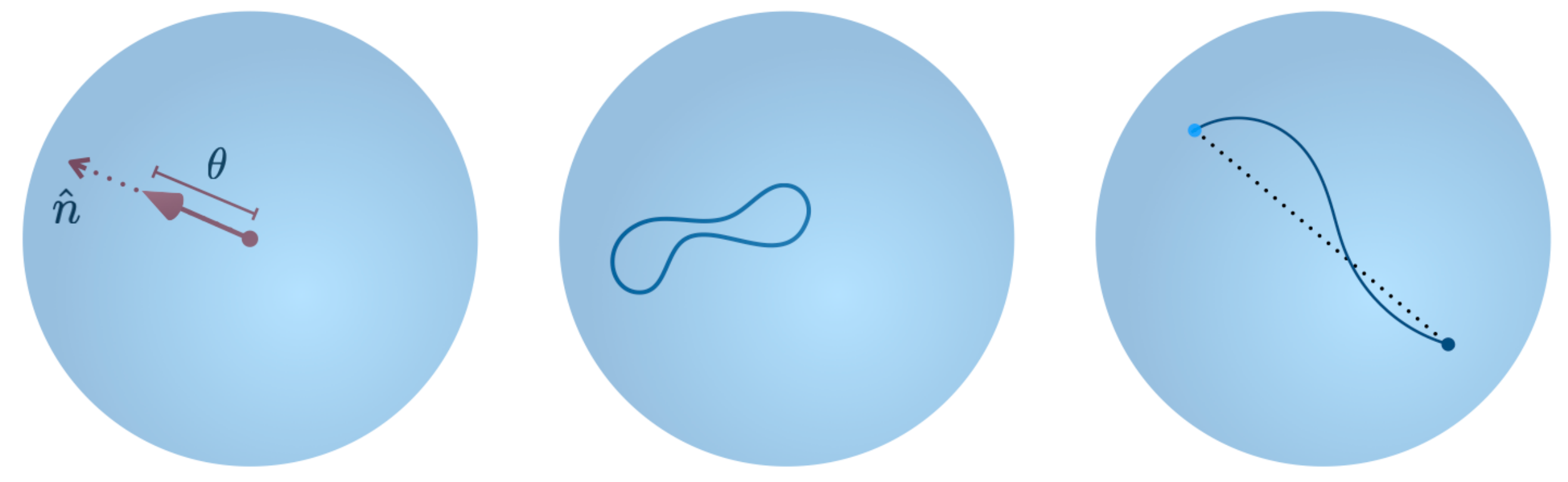}
\caption{Left: $SO(3)$ as a solid sphere. Each point within the sphere corresponds to a unique rotation. The direction from the origin to the point represents the axis of rotation. The distance from the origin to the point represents the angle of rotation. Centre: A `trivial' loop consisting of a closed path that can be smoothly shrunk to a point. Right: A `non-trivial' loop, consisting of a path that connects antipodal points on the surface.}
\label{fig.twoloops}
\end{figure}

The elements of $SO(3)$ can be expressed in various representations\cite{Morrison1987}. We use the axis-angle representation as it best brings out the connectivity of the space. Following Euler's rotation theorem, any rotation in three dimensions can be characterized as $R(\hat{n},\theta)$ using two quantities: an axis $\hat{n}$ (a unit vector) and an angle $\theta$. We can view any given rotation as a vector $\vec{\rho} = \theta ~\hat{n}$, with orientation fixed by the axis and length set by the angle. We can restrict the length of the vectors to $\theta \leq \pi$, using a property of rotations about opposite axes: $R(\hat{n},\pi+\theta) \equiv R(-\hat{n},\pi - \theta)$. With these arguments, we can give a geometric interpretation to $SO(3)$. It corresponds to a solid sphere of radius $\pi$, with each point within the sphere corresponding to a unique rotation operation as depicted in Fig.~\ref{fig.twoloops}(left). However, careful attention must be paid to the surface of this sphere. As $R(\hat{n},\pi) \equiv R(-\hat{n},\pi)$, antipodal points on the surface are identified. That is, each point on the surface is, in fact, the same as its partner at the other end of a diagonal passing through the centre.

We now seek to characterize closed loops within this space\cite{Balakrishnan2018}. We have simple loops as shown in Fig.~\ref{fig.twoloops}(centre). They can be smoothly deformed to a point. We have a second type consisting of loops that connect antipodal points on the surface. While these loops are closed, they cannot be shrunk to a point. We designate these two classes of loops as trivial and non-trivial respectively. A loop of any complexity, e.g., one that passes through several pairs of antipodal points, can be smoothly deformed into one of these classes. 

The $SO(3)$ space is the configuration space of a rigid rotor. To see this, consider an object that is free to rotate, with its centre of mass held fixed. We may designate a particular configuration of the object as our reference, say the configuration at a particular instant of time. The configuration at any other time can be obtained by effecting a rotation, $R (\hat{n},\theta)$, on the reference configuration. In other words, the configuration at any time can be described using a rotation matrix, $R (\hat{n},\theta)$. When this problem is quantized, we arrive at a wavefunction defined on $SO(3)$ space, $\psi(\hat{n},\theta)$. The system is described by the Hamiltonian,
\bea
\hat{H} = \frac{\hat{L}_x^2}{2 I_x} + \frac{\hat{L}_y^2}{2 I_y} + \frac{\hat{L}_z^2}{2 I_z},
\label{eq.Ham}
\eea 
where $\hat{L}_{x/y/z}$ are angular momentum operators defined in the body-fixed frame along the three principal axes. The quantities $I_{x/y/z}$ represent the corresponding moments of inertia. In the rest of this article, in the interest of simplicity, we will consider two cases: (i) $I_x = I_y= I_z \equiv I_0$, known as the `spherical top'. This case arises when the rotating object is a perfect sphere. (ii) $I_x = I_y = I_z/\alpha \equiv I_0$, with $\alpha \neq 1$. This case is known as the `symmetric top', arising in the context of an ellipsoidal rigid body. 

The spectrum of this problem was first worked out by Casimir in 1931. Eigenstates can be labelled by three quantum numbers, $j$, $m$ and $m'$. The first, $j$, represents the total angular momentum. It takes non-negative integer values, $j=0,1,2,3$, etc. The second and third quantum numbers denote angular momenta in the body-fixed frame and in the space-fixed frame respectively\cite{Zare}. They are both defined with respect to an arbitrarily chosen $z$-axis. Each takes one of $(2j+1)$ possible values with $m, m'\in \{-j,-j+1,\ldots,j-1,j\}$. The eigenstates (wavefunctions) are complex functions defined over $SO(3)$ space, given by Wigner $D$ matrices\cite{Varshalovich1988,Edmondsbook,Atkinsbook}, 
\bea
\psi_{j,m,m'} (\hat{n},\theta)= D_{m',m}^j (\hat{n},\theta)^*
\eea
where
\bea
D_{m',m}^j (\hat{n},\theta) &=&  \langle j,m'\vert e^{-i\theta \hat{n}\cdot \vec{\hat{L}} } \vert j,m\rangle.
\label{eq.Dmat}
\eea
Here, $\vec{\hat{L}} = (\hat{L}_x,\hat{L}_y,\hat{L}_z)$ is the angular momentum vector and $\vert j,m\rangle$'s are the usual spherical harmonics. Note that we have expressed the Wigner $D$ matrices in the axis-angle representation here, as opposed to the more-commonly-used Euler angle representation\cite{Rose1957}. In the spherical top rotor, the eigenenergies are given by $\epsilon_j = \frac{\hbar^2}{2I_0} j(j+1)$. As the energies only depend on $j$, each level has a $(2j+1)^2$-fold degeneracy -- corresponding to all possible values of $m$ and $m'$. 
In the symmetric top rotor, eigenenergies are given by $\epsilon_{j,m} = \frac{\hbar^2}{2I_0}\big[ j(j+1) - \gamma m^2 \big] $, where $\gamma= 1 -\alpha $. Each level has a degeneracy of $(2j+1)$, corresponding to different choices for $m'$.

\section{Berry phase structure in the rigid rotor}
\label{sec.berry}

We consider the Berry phase attached to a loop in a generic time-reversal-symmetric system. A loop can be traversed in two directions, which will result in opposite values for the Berry phase. However, two trajectories that correspond to motion in opposite directions are time-reversed copies of one another. In a system with time-reversal symmetry, they must accrue the same Berry phase. With these arguments, the Berry phase must satisfy $\theta_B \equiv -\theta_B$. This has two possible solutions: $\theta_B = 0$ or $\pi$. As these are two discrete values, we argue that Berry phase should be a topological quantity that is invariant under smooth deformations of the path. In particular, if the Berry phase is $\pi$ for a certain path, it must have the same value for all topologically equivalent paths. A trivial path in $SO(3)$, by definition, can be smoothly deformed to a point -- a limiting case with no scope for a Berry phase. We conclude that no trivial path can have an attached Berry phase. 

We next consider non-trivial paths as shown in Fig.~\ref{fig.twoloops}(right). As argued above, time-reversal symmetry constrains the Berry phase of a closed loop to be $0$ or $\pi$. With either value, the Berry phase must be robust to smooth deformations of paths. In $SO(3)$, it can be shown that all non-trivial paths are topologically equivalent\cite{Balakrishnan2018}. That is, any two non-trivial paths can be smoothly deformed into one another. As a result, there are only two possibilities: (i) all trivial and non-trivial paths have a Berry phase of zero. This is the traditional formulation of the rigid rotor problem whose solutions were discussed in Sec.~\ref{sec.rrotor} above. (ii) All trivial paths have Berry phase of zero while all non-trivial paths have a Berry phase of $\pi$. The latter case is the focus of this article. It amounts to a non-trivial boundary condition for the rigid rotor problem. The stationary states are smooth in the interior of the $SO(3)$ sphere. However, they reverse sign under a non-trivial path. This translates to anti-periodic boundary conditions across any diameter of the $SO(3)$ sphere. 

In the context of a rigid rotor, this situation can be described as follows. Consider rotations about an arbitrary axis, $\hat{n}$, denoted as $R(\hat{n},\theta)$. If $\theta$ is taken to run from $-\pi$ to $\pi$, these rotations lie on a diagonal in the $SO(3)$ sphere of Fig.~\ref{fig.twoloops}. As we traverse this non-trivial loop connecting antipodal points, the rigid body effectively rotates by $2\pi$ about a fixed axis. This operation should attach a negative sign to the wavefunction. In the usual problem of rigid body dynamics, such a negative sign does not arise. It can not be easily realized in a physical setup, say by distributing electric charge on the body and threading a magnetic field. However, this situation naturally arises in magnetic analogues of the rigid rotor as we show in the following sections.

To conclude this section, we discuss a simpler problem where the role of a $\pi$-Berry phase can be easily understood. Consider a particle on a circle, a space parametrized by an angle variable, $\phi\in(0,2\pi]$. The non-trivial paths here correspond to a full revolution around the circle. If the circle is threaded by a $\pi$ flux, the particle's wavefunction picks up a negative sign after a revolution. The resulting wavefunctions are of the form $e^{i (n+\frac{1}{2})\phi}$, where $n$ is any integer. These states reverse sign under $\phi \rightarrow \phi+2\pi$, but are periodic under $\phi\rightarrow \phi+4\pi$. They are double-valued, with two possible values for any given $\phi$. These features are closely mirrored by the rigid rotor when a Berry phase of $\pi$ is introduced. 

\section{Drawing solutions from the $SU(2)$ rotor }
\label{sec.sol}
We have established that time-reversal symmetry allows for a situation in which non-trivial paths in $SO(3)$ are associated with a $\pi$-Berry phase. In order to find the corresponding stationary states, we appeal to the $SU(2)$ rotor. A generic $SU(2)$ matrix can be represented using three parameters: a unit vector $\hat{n}$, an angle $\theta\in [0,\pi]$ and an Ising variable $\mu = \pm 1$,
\bea
R_{SU(2)}(\hat{n},\theta,\mu) = \mu ~e^{ i \frac{\theta}{2} \hat{n}\cdot \vec{\sigma}},
\label{eq.su2def}
\eea
where $\vec{\sigma} = (\sigma_x,\sigma_y,\sigma_z)$ is a vector of Pauli matrices. This form brings out the relation between $SU(2)$ and $SO(3)$ spaces\cite{Wignerbook,Edmondsbook}. They bear a two-to-one relation, with $SU(2)$ consisting of two copies of the $SO(3)$ solid sphere -- one for each value of $\mu$. The two solid spheres of $SU(2)$ share a common boundary, with 
\bea
R_{SU(2)}(\hat{n},\pi,+1) = R_{SU(2)}(-\hat{n},\pi,-1),
\eea
which follows from the definition in Eq.~\ref{eq.su2def}.

The $SU(2)$ rotor is very similar to the $SO(3)$ rigid rotor, with a Hamiltonian of the same form as in Eq.~\ref{eq.Ham}. Its stationary states are also similar, given by Wigner $D$ matrices of  Eq.~\ref{eq.Dmat} above. However, in the $SU(2)$ case, the total angular momentum quantum number may take non-negative values that are integers or half-integers. That is, $j=0, \frac{1}{2}, 1, \frac{3}{2},2,$ etc. For each $j$, the $m$ and $m'$ quantum numbers take values from $\{-j,-j+1,\ldots,j-1,j\}$. The stationary state wavefunctions are also similar to the $SO(3)$ case. However, careful attention must be paid to continuity at the boundary of the sphere, as we discuss below. 

The Wigner $D$ matrices are complex-valued functions of $\hat{n}$ and $\theta$. However, $SU(2)$ space has an additional coordinate in $\mu$. In order to define wavefunctions in a consistent fashion, we note that the Wigner $D$ matrices satisfy 
\bea
\psi(-\hat{n},\pi) =(-1)^{2j}\psi(\hat{n},\pi).
\label{eq.jrelation}
\eea
This relation, demonstrated explicitly in App.~\ref{app.periodicity}, connects the values of the wavefunction at antipodal points of the sphere. We see that stationary states fall into two classes, based on the value of $j$. Solutions with integer values ($j=0,1,2,3,\ldots$) have the same value at any pair of antipodal points. Those with half-integer values ($j=\frac{1}{2},\frac{3}{2},\frac{5}{2},\ldots$) have a relative minus sign. To define wavefunctions that are smooth over the $SU(2)$ configuration space, we take
\bea
\psi_{SU(2)}(\hat{n},\theta,\mu) = \left\{ 
\begin{array}{c}
D_{m',m}^j (\hat{n},\theta)^*,~~j=0,1,2,\ldots \\
\mu ~D_{m',m}^j (\hat{n},\theta)^*,~~j=\frac{1}{2},\frac{3}{2},\ldots
\end{array}\right..
\eea 
For integer-valued $j$'s, the wavefunctions are identical between the two spheres. For half-integer $j$'s, the wavefunctions differ by a negative sign. These forms lead to smooth evolution across the boundary that separates the two spheres of $SU(2)$. In the language of Ref.~\onlinecite{Wignerbook}, integer and half-integer $j$ values correspond to even and odd representations respectively.

We now revert to the problem of the $SO(3)$ rigid rotor with a $\pi$-Berry phase. We seek to find eigenstates of the Hamiltonian in Eq.~\ref{eq.Ham} that are smooth within the $SO(3)$ sphere and \textit{anti-periodic} across diagonals. These requirements are precisely met in the $SU(2)$ solutions with half-integer $j$ values, as seen from Eq.~\ref{eq.jrelation}. We conclude that the required stationary states are Wigner $D$ matrices with $j=\frac{1}{2}, \frac{3}{2}, \frac{5}{2}$, etc. The expressions for the eigenenergies and level degeneracies are the same as those given in Sec.~\ref{sec.rrotor}, but with $j$ taking half-integer values. In Tab.~\ref{tab.spec}, we describe the resulting spectrum and compare it with that of the traditional Berry-phase-free formulation.

\begin{table*}
\begin{tabular}{|c|c|c|c|c|}
\hline
Berry phase  & Total angular momentum quantum number & $m$ and $m'$ quantum numbers & Eigenvalue & Degeneracy\\
\hline
$\begin{array}{c}
0 \\
\pi \end{array}$
& 
$\begin{array}{c}
j = 0,1,2,3,\ldots \\
j = \frac{1}{2},\frac{3}{2},\frac{5}{2},\ldots
\end{array}$
& 
$m / m' = -j, -j+1,\ldots,j-1,j$
&
 $\epsilon_j = j(j+1)$ & $(2j+1)^2$ \\
\hline
\end{tabular}
\caption{Spectrum of the spherical-top rigid rotor with and without Berry phase.}
\label{tab.spec}
\end{table*}

\section{Rigid rotor as an effective description: the triangle antiferromagnet}
The rigid rotor can emerge as the low energy description of certain quantum magnets. Our discussion below follows a general principle laid down in Ref.~\onlinecite{Khatua2019}. The physics of a quantum magnet, at low energies, resembles that of a single particle problem. The particle moves in the abstract space of all classical ground states. This mapping is readily seen from the low-lying portion of the magnet's energy spectrum. Below, we will consider a family of quantum magnets where the set of classical ground states is isomorphic to $SO(3)$. Their low-energy physics corresponds to a particle moving in $SO(3)$ -- a rigid rotor.

We begin with perhaps the simplest example -- a three-spin magnet with spin-$S$ moments at the corners of an equilateral triangle. Neighbouring spins interact via an antiferromagnetic Heisenberg coupling. The Hamiltonian for this system can be written as
\bea
H = J\sum_{(jk)} \vec{\hat{S}}_{j} \cdot \vec{\hat{S}}_{k} \sim  \frac{J}{2}\big(\sum_j\vec{\hat{S}}\big)^2,
\eea
where the sum over $(jk)$ runs over all pairs of spins. This can
be rewritten as the square of a sum over each spin. We have expressed the Hamiltonian as the square of the total spin, by removing a constant term. In the classical limit, the spins can be viewed as three-component vectors. The classical energy is minimized when the three spin vectors add to zero. This can only happen if all three spins lie on a plane and form the sides of an equilateral triangle. As is typical in frustrated systems, there are, in fact, many classical ground states. The choice of the ordering plane corresponds to a choice of normal unit vector, $\hat{n}$. Within the plane, the first spin may be oriented in any direction. This corresponds to an angle variable, $\theta$. These two parameters, $\hat{n}$ and $\theta$, suggest that the space of classical ground states is equivalent to $SO(3)$ which is also parametrized by a unit vector and an angle. Indeed, it can be rigorously shown that any classical ground state can be obtained from a reference ground state, by effecting a global $SO(3)$ rotation. We say that the set of classical ground states is $SO(3)$, i.e., the set of classical ground states has a one-to-one and onto mapping with $SO(3)$.  

\subsection{Low-energy effective theory }
\label{ssec.tri_eff}
The low energy physics of this magnet can be systematically studied using a non-linear sigma model approach. A detailed calculation, applicable to an entire family of magnets, is outlined in Sec.~\ref{ssec.nlsm} below. Here, we describe the final result for the case of the triangle magnet. The partition function can be expressed within the spin path integral formalism. It involves an integral over all paths in configuration space, 
\bea
\mathcal{Z} &=&\int \big\{\prod_{j=1}^3 \mathcal{D}S_j \big\} e^{-\mathcal{L}} = \sum_{\mathrm{loops}} e^{-\mathcal{L}_{loop}},\\
\mathcal{L}_{loop} &=& i{\Theta}_{loop} + J\int_0^\beta d\tau \sum_{(jk)} \vec{S}_j (\tau) \cdot \vec{S}_k (\tau).
\label{eq.path_int_action}
\eea
The loops lie within the space of all classical configurations of spins. They run over imaginary time, from $\tau=0$ to $\beta$, where $\beta$ is the inverse temperature. Each loop contributes with an action that has two terms, the Berry phase, $\Theta_{loop}$, and the energy. The latter can be written as an expansion in powers of $S$, with the leading $\mathcal{O}(S^2)$ term corresponding to the classical energy. 

In the large-$S$ low-energy regime, we may restrict the path integral to configurations that are close to classical ground states. Any `hard' deviation (one that increases the classical energy) is exponentially suppressed. In this regime, we may restrict our attention to loops that can be decomposed into two pieces: (i) a closed loop within the space of classical ground states, and (ii) a small fluctuation out of the ground state space. The former determines the topological character of the loop. Here, the space of classical ground states is $SO(3)$. As a result, a loop can be classified as trivial (shrinkable to a point) or non-trivial, as shown in Fig.~\ref{fig.twoloops}. The fluctuation out of the ground state space represents a small deformation that does not modify topological character. 

After several simplifications, the action for a given loop takes a remarkably simple form,
\bea
\mathcal{L}_{loop} = 6\pi i S \nu -i \vec{L}'\cdot \vec{V} + \beta_3 \vec{L}'^2,
\eea  
Here, $\nu=0,1$ is a topological index. It is $0$ for a trivial path within $SO(3)$ and $1$ for a non-trivial path. The remaining terms have a standard form -- the action of a spherical top rigid rotor. Here, $\vec{L}'$ represents the angular momentum, $\vec{V}$ represents angular velocity and $\beta_3$ is an energy scale, proportional to $J$. Explicit expressions for these quantities are given in Sec.~\ref{ssec.nlsm} below.

We have arrived at a remarkable picture. The effective theory for a triangle antiferromagnet is precisely a spherical top rigid rotor. This result is well known and has been used in field theoretic studies of triangular lattice antiferromagnets, where each triangular motif is viewed as a rigid rotor\cite{Dombre1989,Azaria1993,Diptiman1993}. However, the Berry phase term has not been adequately appreciated in earlier studies. Our analysis shows that it can play a strong role. For integer $S$, the Berry phase is inconsequential as it is an integer multiple of $2\pi$. This leads to an effective theory of a traditional spherical top rotor. However, for half-integer values of $S$, a Berry phase of $\pi$ is attached to non-trivial paths in $SO(3)$. The effective theory is the spherical top rigid rotor with a $\pi$-Berry phase, precisely as described in Secs.~\ref{sec.berry} and \ref{sec.sol} above.  

\subsection{Rigid rotor in the quantum spectrum}
\label{ssec.tri_spectrum}
We now discuss the energy spectrum of the triangle antiferromagnet. As the Hamiltonian is proportional to the total-spin-squared, its eigenvalues are $\frac{J}{2} S_t (S_t+1)$, where $S_t$ is the total spin. We arrive at a simple problem of angular momentum addition,
\bea
 S\otimes S\otimes S = \{0 \oplus1 \oplus 2\ldots \oplus 2S\}\otimes S.
\eea
We have added the first two spins before adding the third. The final result depends on the nature of $S$. 

We first consider integer values of $S$, where we find $S_t \in \{0,1,2,\ldots\}$. The ground state, corresponding to $S_t=0$, is non-degenerate. The value $S_t=0$ only arises in the case where the sum of the first two spins is $S$. The first excited state corresponds to $S_t = 1$, which occurs when the first two spins add to $S-1$, $S$ or $S+1$. In each case, as $S_t = 1$ is a triplet, we have an additional three-fold degeneracy. This results in a net nine-fold degeneracy of the first excited state. Higher energy levels can be found from similar arguments. We summarize as follows: eigenstates correspond to $S_t=0,1,2,\ldots$ with each level having a degeneracy of $(2 S_t +1)^2$. Note that this only represents the low-energy spectrum -- the pattern holds for $S_t \leq S$. Remarkably, the low-energy spectrum is precisely that of a spherical top rigid rotor with no Berry phase, as described in Sec.~\ref{sec.rrotor}.

\begin{figure}
\includegraphics[width=3.4in]{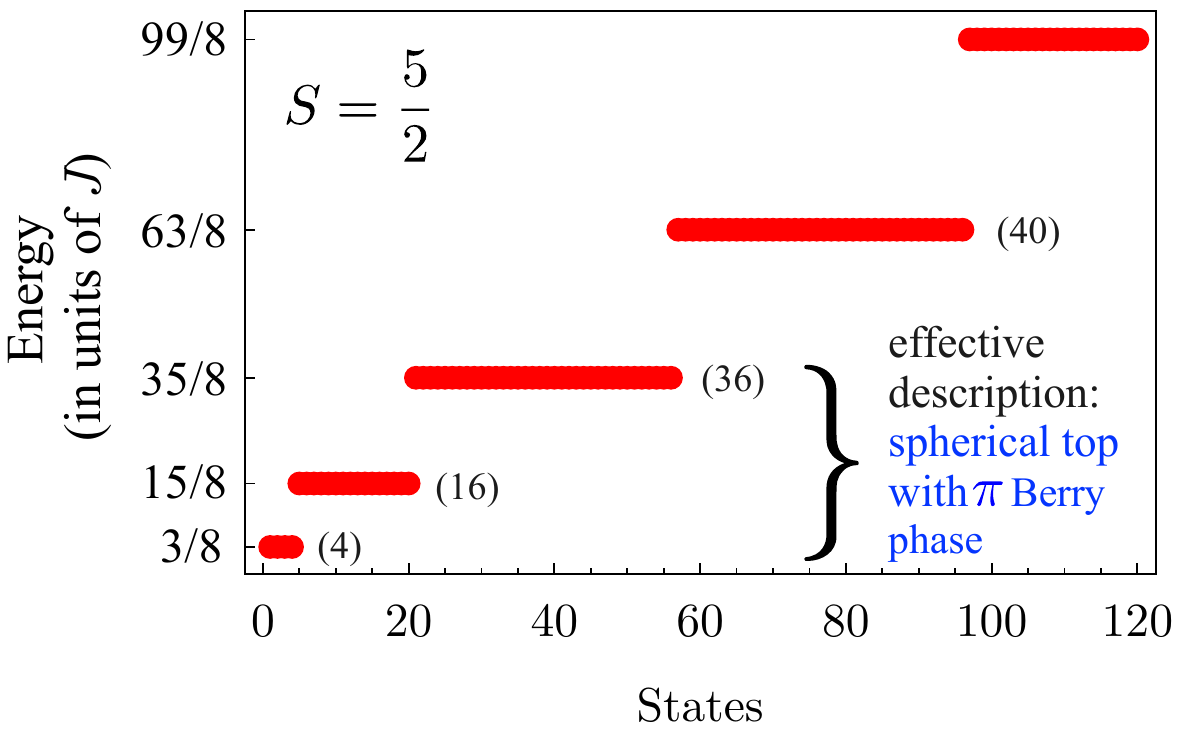}
\caption{The spectrum of a triangle antiferromagnet with a half-integer value of $S$. The degeneracy of each level is shown in parentheses. Within the low energy window indicated, the spectrum is quantitatively equivalent to that of a spherical top rigid rotor with a $\pi$-Berry phase.}
\label{fig.trispec}
\end{figure}

For half-integer values of $S$, the spectrum is qualitatively different as we have $S_t = \frac{1}{2},\frac{3}{2},\frac{5}{2},$ etc. The ground state corresponds to $S_t=\frac{1}{2}$, which can occur when the first two spins add to $S\pm \frac{1}{2}$. In addition, each $S_t=\frac{1}{2}$ level has an inherent two-fold degeneracy. This leads to a net four-fold ground state degeneracy. On the same lines, the first excited state corresponds $S_t =\frac{3}{2}$ and is sixteen-fold degenerate. In summary, eigenstates are labelled by $S_t = \frac{1}{2},\frac{3}{2},\frac{5}{2},\ldots$, with degeneracies $(2S_t+1)^2$. This picture holds for $S_t \leq S$, representing the low-energy spectrum. Remarkably, this is the spectrum of a spherical top rigid rotor with a Berry phase of $\pi$. The spectrum for the case of $S=5/2$ is plotted in Fig.~\ref{fig.trispec}. The effective description in terms of a spherical top with a $\pi$-Berry phase holds at low energies, for $E\lesssim \frac{35J}{8}$.

The energy spectra, obtained from analytical arguments, are consistent with the low-energy theory outlined in Sec.~\ref{ssec.tri_eff} above. The triangular antiferromagnet, at low energies, is a realization of the spherical top rigid rotor. Depending on the value of $S$, the rotor may have a Berry phase of zero or $\pi$.   

\section{Odd-polygon antiferromagnets}
\label{sec.oddgon_antiferromagnet}
We next consider the more general case of odd-polygon antiferromagnets. We take polygons with $N$ vertices, where $N=3,5,7,9,$ etc. We have spin-$S$ moments at each vertex with Heisenberg antiferromangetic couplings between nearest neighbours,
\bea
H = J\sum_{i=1}^N \vec{\hat{S}}_{i} \cdot \vec{\hat{S}}_{i+1},
\label{eq.Haml}
\eea
where $\vec{\hat{S}}_{N+1}\equiv \vec{\hat{S}}_{1}$. Unlike the triangle antiferromagnet, this Hamiltonian cannot be written as the square of the total spin. Nevertheless, as we show below, its structure and high degree of symmetry lead to an elegant low-energy description. 

We first discuss the classical ground states of this problem. For polygons with even $N$, there is no frustration. The classical ground state is a N\'eel antiferromagnet at the ordering wavevector $\pi$. Moments alternate between two opposite orientations as we move from one site to the next, reverting to the initial orientation when we return to the starting point. However, when $N$ is odd, N\'eel order cannot be accommodated. Energy is minimized by ordering at $\pi \pm \pi/N$. These are the closest wavevectors to $\pi$ that leave the system invariant after $N$ unit translations. Effectively, the classical ground state is a coplanar state with the ordering-plane chosen spontaneously. Within the plane, neighbouring spins subtend an angle $\pi \pm \pi/N$ with one another. When moving from one site to the next clockwise, the subtended angle is always the same, either $\pi + \pi/N$ or $\pi - \pi/N$. A rigorous discussion can be found in Ref.~\onlinecite{Schmidt2003}. 

Furthermore, all classical ground states can be reached from an arbitrary reference state by effecting a global spin rotation. The choice of ordering plane corresponds to choosing a normal unit vector $\hat{n}$. We may fix its orientation to be parallel to $\vec{S}_2 \times \vec{S}_1$. Within the plane, the orientation of the first spin corresponds to choosing an angle $\theta$. In order to avoid double-counting, $\theta$ can be restricted from 0 to $\pi$ while $\hat{n}$ ranges over all orientations. This parametrization is equivalent to specifying an $SO(3)$ rotation using the axis-angle representation. We assert that the set of all classical ground states is isomorphic to $SO(3)$. 

\begin{figure}
\includegraphics[width=3.3in]{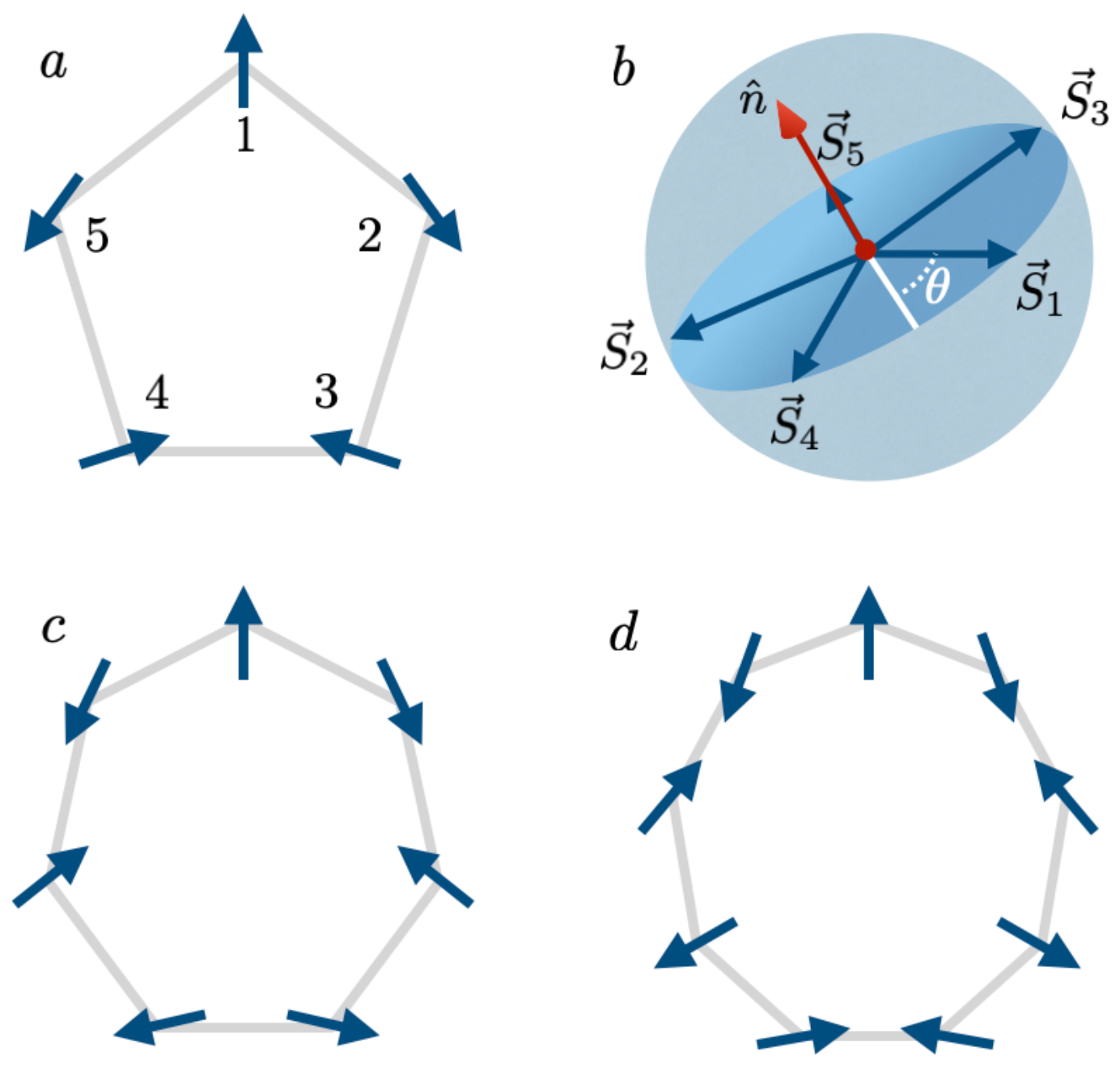}
\caption{Ordering in odd antiferromagnetic polygons. (a) A reference classical ground state on the pentagon. All spins lie on the same plane with neighbouring spins subtending an angle of $4\pi/5$ with one another. (b) Any classical ground state can be obtained by rotating the reference state. The orientations of the five spins are shown. The plane of ordering is fixed by the normal vector $\hat{n}$ while the position of the first spin is set by $\theta$, measured from an arbitrary axis. All ground states can be obtained by varying the normal vector $\hat{n}$ or the angle $\theta$. (c) A reference ground state on a heptagon ($N=7$). (d) A reference ground state on a nonagon ($N=9$).  }
\label{fig.oddorder}
\end{figure}

Below, we discuss a low-energy effective theory for odd-$N$ polygons. For $N=3$, our theory yields a spherical top rotor while $N>3$ gives rise to a symmetric top. Subsequently, we present the energy spectrum obtained from exact diagonalization.  

\subsection{Effective theory}
\label{ssec.nlsm} 
We derive a low-energy effective description using the non-linear sigma model approach. Our approach applies to the entire family of odd-polygon antiferromagnets, where $N=3,5,7,9,$ etc. For any $N$, the classical ground states are coplanar and accessible by global rotations acting on a reference ground state as shown in Fig.~\ref{fig.oddorder}. We define a reference state in the $x-y$ plane, 
\bea
\nonumber \{\vec{S}_1, \vec{S}_2, \cdots,\vec{S}_N \}_{ref.} \equiv  S\{\hat{n}_1, \hat{n}_{2},\ldots, \hat{n}_{N-1}, \hat{n}_{N} \} \\
= S\{\hat{\nu}_\phi, \hat{\nu}_{\phi+\theta_N}, \hat{\nu}_{\phi+2 \theta_N}, \ldots, \hat{\nu}_{\phi+(N-1)\theta_N} \},
\label{eq.Sref}
\eea 
where $\hat{\nu}_\xi = \cos \xi ~\hat{x} + \sin \xi~ \hat{y}$ and $\theta_N = \pi + \pi/N$. The orientation of the first spin is fixed by an angle $\phi$ ranging from $0$ to $2\pi$. A generic classical ground state can now be written as $\{\vec{S}_1, \vec{S}_2, \cdots,\vec{S}_N \} = SR(\hat{n},\theta)\{\hat{n}_1,\hat{n}_2,\cdots,\hat{n}_N\}$, where the rotation $R(\hat{n},\theta)$ acts on each of the $N$ unit vectors.

In the spirit of a low energy theory, we introduce a small deviation away from the ground state. The deviation is parametrized by a vector, $\vec{L}$, which will turn out to be proportional to uniform polarization,
\begin{eqnarray}
\label{eq.polygon_fluc}
&&\vec{S}_i = S\hat{\Omega}_i=\frac{S R(\hat{n},\theta) \big\{ \hat{n}_i + M_i \vec{L}/S \big\}}{\sqrt{1+ (M_i\vec{L}/S)^2}}\nonumber\\
&\approx & S R(\hat{n},\theta) \left\{ \hat{n}_i \left(1-\frac{\vec{L}\cdot M_i\vec{L}}{2S^2}\right)+ M_i \vec{L}/S \right\},
\end{eqnarray}
where $M_i^{ab} = (\delta^{ab} -\hat{n}_i^a \hat{n}_i^b )$ is a tensor operator that projects onto the plane perpendicular to $\hat{n}_i$. When it acts on a vector to the right, it picks out the component perpendicular $\hat{n}_i$. Note that $M_i^2 = M_i$. In this parametrization, $\hat{n}$ and $\theta$ represent `soft' modes that preserve the classical energy. The vector, $\vec{L}$, encodes modes that are canonically conjugate to the soft modes, as we will show below. It represents a deviation that cants each spin towards the direction of $\vec{L}$. 

Apart from $\vec{L}$, we may introduce many other deformations in Eq.~\ref{eq.polygon_fluc}, e.g., a staggered canting along each direction. However, such fluctuations can be integrated out from the path integral action. We are interested in the low-energy physics involving the soft modes and their conjugate degrees of freedom.  

We next describe the magnet using the well-known spin-path-integral formalism\cite{Auerbach,Fradkin}. The partition function is written as a sum over loops in the space of all classical configurations. Each loop is associated with an action that has two terms: the Berry phase and the energy. We evaluate these contributions using the low energy spin-parametrization of Eq.~\ref{eq.polygon_fluc}.\\
{\bf{Berry phase:}} This represents a geometric contribution, proportional to the sum of solid angles swept out by each spin. It can be written as
\bea
iS\int_0^\beta d\tau \sum_{j=1}^N\vec{A}(\hat{\Omega}_j)\cdot\partial_\tau(\hat{\Omega}_j),
\eea
where $\vec{A}(\hat{\Omega}_j)$ is the vector potential of magnetic monopole at the origin. With the parametrization of Eq.~\ref{eq.polygon_fluc}, the Berry phase takes the form
\bea
\nonumber iS\int_0^\beta d\tau \sum_{j=1}^N\vec{A}\left(R \big\{ \hat{n}_j + M_j \frac{\vec{L}}{S} \big\}\right)\\
 \cdot\partial_\tau\left(R \big\{ \hat{n}_j + M_j \frac{\vec{L}}{S} \big\}\right) 
+ \mathcal{O}\Big(\frac{1}{S}\Big).
\label{eq.Berryexp}
\eea
We retain terms of $\mathcal{O}(S^0)$, in the spirit of an expansion in powers of $S$. As we operate in the low-energy regime, we take each trajectory to consist of two parts: a closed loop entirely within the classical ground state space and a small deviation out of it. The latter amounts to a small, smooth deformation of the former. This picture, after some straightforward simplifications, leads to a remarkably simple form,
\bea
2\pi i N S\nu - i\vec{L}^{'}\cdot\vec{V} + \mathcal{O}\Big(\frac{1}{S}\Big).
\label{eq.Berryform}
\eea  
The leading $\mathcal{O}(S)$ term only depends on the trajectory within the classical ground state space. This gives a topological contribution, with $\nu=0$ for contractible loops and $\nu=1$ for non-trivial loops. This is explicitly demonstrated in App.~\ref{app.Berry_phase}. 

In writing the $\mathcal{O}(S^0)$ term, we have defined two new variables. The first is $\vec{L}^{'} = \sum_{j=1}^N \vec{S}_j = RM\vec{L}$, where $M$ is given by $\sum_{j=1}^N M_j = \mathrm{Diag} \{N/2,N/2,N\}$.  The vector $\vec{L}^{'}$ has a clear physical interpretation, as the net magnetization of the system. 
The second variable is $V_\alpha = -\frac{1}{2}\epsilon_{\alpha\beta\gamma}\{(\partial_\tau R)R^{-1}\}^{\beta\gamma}$. If we take $R$ to denote the configuration of a rigid body, $V_\alpha$ represents its angular velocity. From the linear coupling between $\vec{L}^{'}$ and the angular velocity, we see that $\vec{L}^{'}$ is canonically conjugate to $R$.

{\bf{Energy:}} Using the parametrization of Eq.~\ref{eq.polygon_fluc} in the Hamiltonian of Eq.~\ref{eq.Haml}, we obtain
\bea
 \beta E_{CGS}  
 + \int_0^\beta d\tau E_2.
\label{eq.oddgon_energy}
\eea
The leading $\mathcal{O}(S^2)$ contribution is $E_{CGS}$, the classical ground state energy. We have $E_{cgs} = NJS^2\cos \theta_N$, where $\theta_N$ is as defined below Eq.~\ref{eq.Sref}. The $\mathcal{O}(S)$ contribution to the energy vanishes as it is linear in $\vec{L}$. As we are expanding about an extremum of the classical energy, all linear terms vanish. The second term in Eq.~\ref{eq.oddgon_energy} is $\mathcal{O}(S^0)$, or equivalently $\mathcal{O}(L^2)$. It is given by 
 \bea
 E_2 &=& J\left[NL^2 - 2\sum_{j=1}^N (\hat{n}_i\cdot \vec{L})^2- \cos(\theta_N)(M\vec{L}\cdot\vec{L}) \right.\nonumber\\
 &&\left. \hspace{0.3cm}+ \cos(\theta_N)\sum_{j=1}^N (\hat{n}_j\cdot \vec{L})(\hat{n}_{j+1}\cdot \vec{L})\right]\nonumber\\
 &=& \beta_N\vec{L}^{'2} - \gamma_N (M\vec{L})_z^{2} = \beta_N\vec{\tilde{L}}^{2} - \gamma_N {\tilde{L}}_z {}^{2} . 
 \label{eq.efftheory}
 \eea
 where the explicit forms for $\beta_N$ and $\gamma_N$ are given in App.~\ref{app.exps}. We have defined $\vec{\tilde{L}} \equiv M \vec{L}$ as the net magnetization in the body-fixed frame.

Putting together the Berry phase and energy terms, the action for a given loop takes the form
 \bea
 \mathcal{L}_{\mathrm{loop}} \approx  2\pi i NS\nu - i\vec{L}^{'}.\vec{V} + \beta_N\vec{\tilde{L}}^{2}- \gamma_N {\tilde{L}}_z^2. 
 \label{eq.sym_top}
 \eea
We have neglected a constant contribution given by $\beta E_{CGS}$, as well as $\mathcal{O}(\frac{1}{S})$ contributions. The result is a recognizable form -- the action of a symmetric top rigid rotor with an additional Berry phase term. For integer values of $S$, the Berry phase can be neglected as it is always an integer multiple of $2\pi$. For half-integer $S$ values, a $\pi$-Berry phase is attached to non-trivial loops. For $N=3$, it turns out that the asymmetry coefficient vanishes ($\gamma_3=0$). This is a special case where we obtain a spherical top rotor. 

The effective action above has characteristic symmetries. It is invariant under rotations of the space-fixed frame. This corresponds to a transformation $R (\hat{n},\theta) \rightarrow  R_0 R(\hat{n},\theta)$ in the parametrization of Eq.~\ref{eq.polygon_fluc}, where $R_0$ is a constant $SO(3)$ matrix. In contrast, a rotation of the body-fixed frame corresponds to modifying the choice of reference state in Eq.~\ref{eq.Sref}. We immediately see that the action is independent of the angle $\phi$ used in Eq.~\ref{eq.Sref}. A generic body-frame rotation may also change the plane of ordering. This only modifies the last term in the action, with ${\tilde{L}}_z^2 \rightarrow {\tilde{L}}_n^2$, where ${\tilde{L}}_n$ is the component normal to the plane of the reference state. 

The action of Eq.~\ref{eq.sym_top} corresponds to characteristic patterns in the energy spectrum. The level-spacing and degeneracies depend on the relative strengths of $\beta_N$ and $\gamma_N$ coefficients. In App.~\ref{app.exps}, we discuss the variation of these parameters with $N$. For the limiting value of $N=3$, $\gamma_3$ vanishes. This leads to the spectrum of a spherical top rigid rotor, with eigenvalues $\beta_3 j(j+1)$. When $S$ is a half-integer, $j=\frac{1}{2},\frac{3}{2},\frac{5}{2},\ldots$, as discussed in Sec.~\ref{ssec.tri_eff} above. For all higher $N$, energy eigenvalues are given by $\beta_N j (j+1) - \gamma_N m^2$. As long as $S$ is a half-integer, $j=\frac{1}{2},\frac{3}{2},\frac{5}{2},\ldots$ with $m = -j,\ldots,j$. Each $(j,m)$ level has a $(2j+1)$-degeneracy, arising from all allowed values of the $m'$ quantum number. For all $N>3$, we find a small, positive value for the asymmetry coefficient $\gamma_N$, with $\gamma_N \lesssim \beta_N/2$. In this regime, the ground state corresponds to $(j=\frac{1}{2}, m =\pm \frac{1}{2})$ with four-fold degeneracy; the first excited state corresponds to $(j=\frac{3}{2}, m=\pm \frac{3}{2})$ with eight-fold degeneracy, and so on. The effective theory yields the same pattern in the energy spectrum for all $N>3$. 

In the thermodynamic limit, where $N\rightarrow \infty$, the effective theory retains its symmetric-top-rotor character. The asymmetry is small with $\gamma_N /  \beta_N \rightarrow \frac{1}{2}$. However, both $\beta_N$ and $\gamma_N$ decrease as $\sim1/N$. This can be interpreted as stiffening of the rotor with the moment of inertia increasing linearly with $N$.

\subsection{Energy spectra using exact diagonalization}
We have argued that odd-polygon antiferromagnets, at low energies, acquire an emergent description in terms of symmetric-top rigid rotors. We now support this assertion with evidence from exact diagonalization spectra. 

{\bf{Methodology:}} We discuss the $N$-site antiferromagnetic chain, where $N$ is odd. The spectrum of the $N=3$ problem can be solved analytically, as presented in Sec.~\ref{ssec.tri_spectrum} above. Here, we discuss a numerical approach for $N=5,7,9$, for various $S$ values. 
The Hilbert space is $(2S+1)^N$-dimensional, growing rapidly with $N$ and $S$. For larger values of $N$ and $S$, we use the following four symmetries to diagonalize the Hamiltonian: a) Invariance under global spin rotations about the $z$ axis. This divides the Hilbert space into orthogonal sectors labelled by $S_z^{total}$. b) Invariance under a global spin rotation by $\pi$ about the spin-$x$ axis. This operation takes $S_z^{total}\rightarrow -S_z^{total}$. As this is a symmetry, we only need to solve the problem in sectors with positive values of $S_z^{total}$. c) Invariance under a unit translation in real space, i.e., $\vec{\hat{S}}_1\rightarrow\vec{\hat{S}}_2\rightarrow\vec{\hat{S}}_3\rightarrow\cdots\rightarrow\vec{\hat{S}}_N\rightarrow\vec{\hat{S}}_1$. This symmetry allows us to subdivide the Hilbert space into blocks of different momenta, $k = 2p\pi/N$ where $p = 0,1,2,\cdots,N-1$. d) Invariance under a mirror operation which transforms sites as $(1,2,\ldots,N-1,N)\rightarrow (N-1,N-2,\ldots,2,1,N)$. This symmetry relates $k=2p \pi/N$ and $k=2(N-p) \pi/N$ subsectors within a given $S_z^{total}$ sector. For $k \neq 0$, this reduces the computational effort by half.

For smaller values for $N$ and $S$ (e.g., for $N=5$, $S\leq 9$), we diagonalize each block of the Hamiltonian matrix to find the spectrum. For larger $S$, block sizes are too large for full diagonalization. However, the Hamiltonian is sparse as it consists only of nearest-neighbour couplings. This allows us to use Lanczos diagonalizatoin implemented using the ARPACKPP package\cite{arpackpp}, focussing on the lowest few eigenvalues.   

\begin{figure}
\includegraphics[height = 5.3 cm]{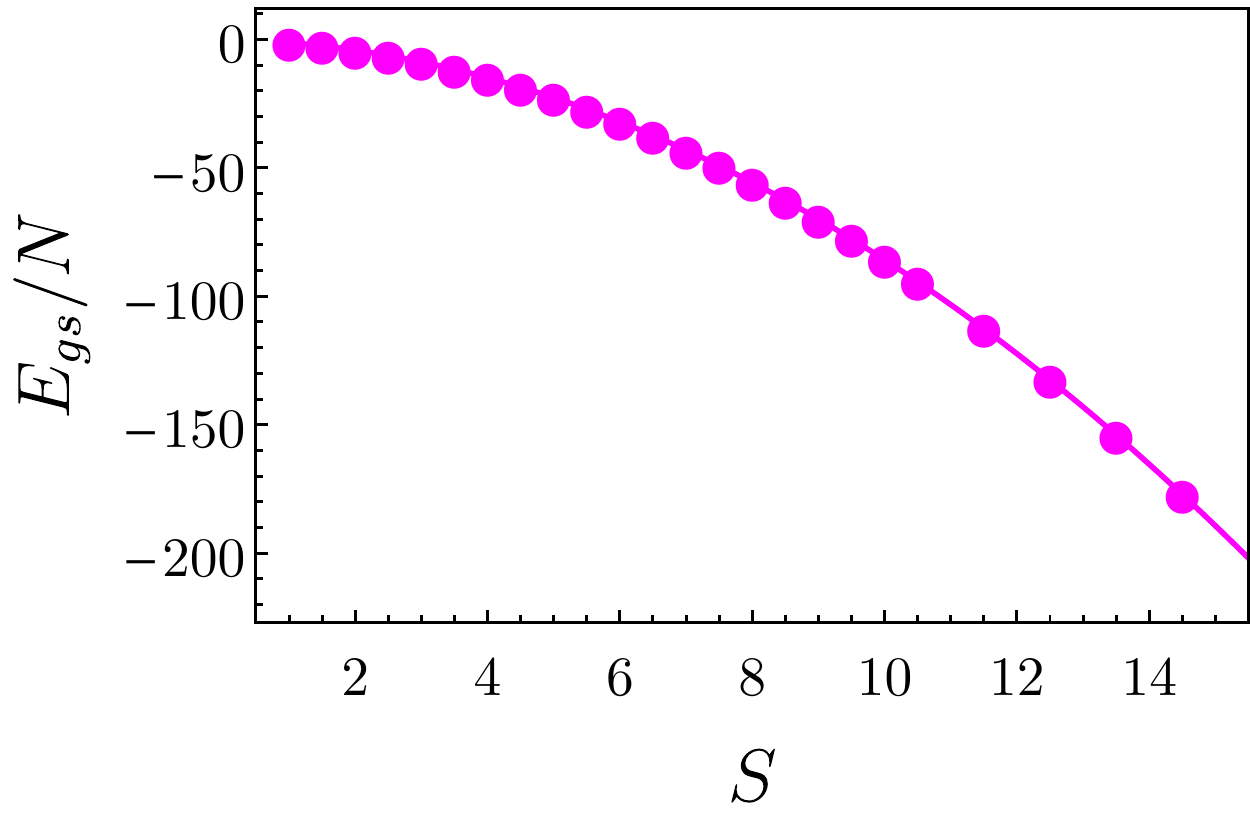}\\
\includegraphics[height = 5.3 cm]{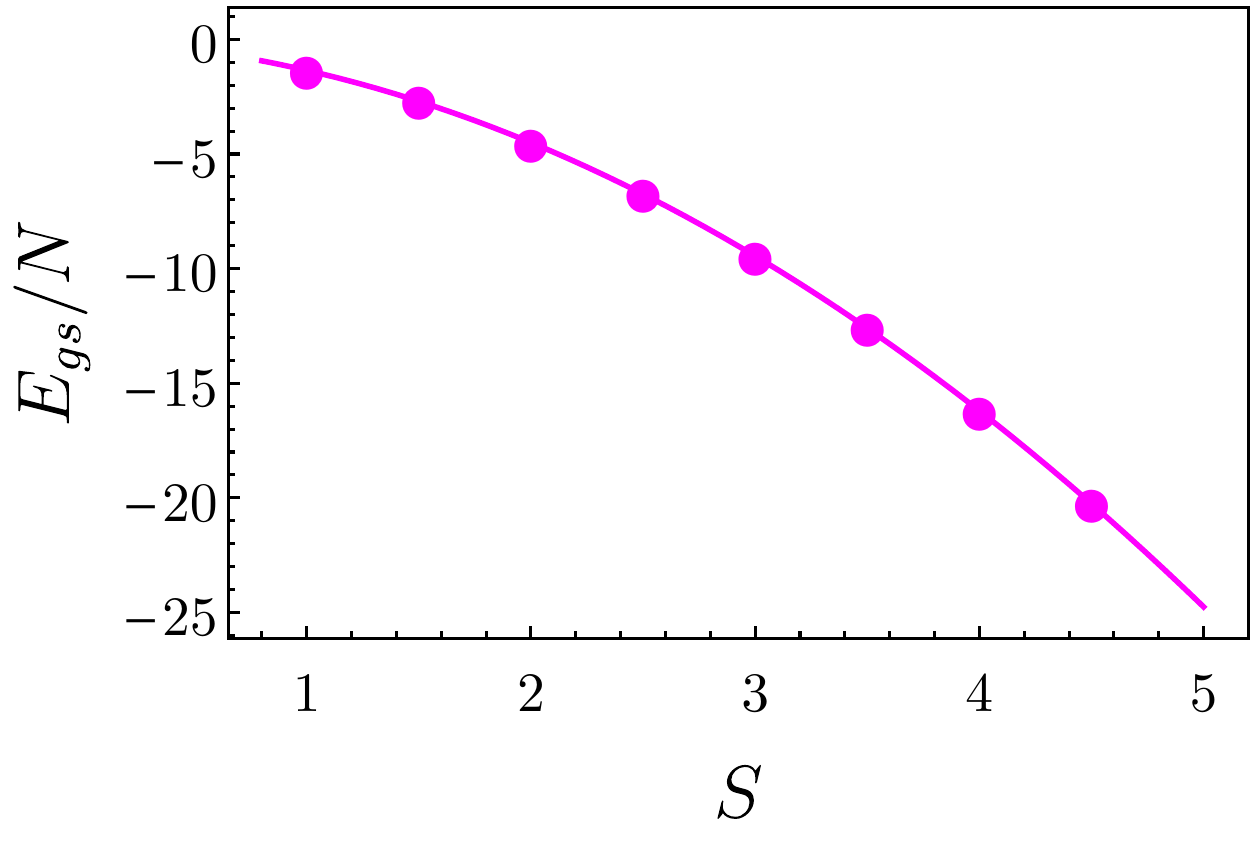}
\caption{Top: Ground state energy per bond in the pentagon vs. spin $(S)$. Datapoints are fit to the curve, $f(S) = 0.0347328 - 0.496368S - 0.808579 S^2$. Bottom: Ground state energy per bond in the hepatgon vs. spin $(S)$. Datapoints are fit to $g(S) = -0.0019346 - 0.448329S - 0.900543S^2$. Energies are measured in units of $J$.}
\label{fig.ground_state_energy}
\end{figure}

{\bf{Ground state energy:}} We first discuss the ground state energy, obtained by numerical diagonalization of the Hamiltonian. Fig.~\ref{fig.ground_state_energy} plots the ground state energy per bond ($E_{gs}/N$) vs. $S$ for the pentagon $(N=5$) and the heptagon $(N=7$). The points fall on smooth curves of the form $E_{gs}/N\sim a S^2 + b S +c$, where $a$, $b$ and $c$ are obtained as fitting parameters. We interpret the $\mathcal{O}(S^2)$ term as the classical energy, while the others are quantum corrections. From fitting the data, the $\mathcal{O}(S^2)$ terms comes out to be $-0.808579 J S^2$ for the pentagon and $-0.900543 J S^2$ for the heptagon. These estimates are consistent with our semiclassical analysis to find the effective theory. The starting point of our analysis was the assertion that the classical ground states are coplanar, with neighbouring spins subtending an angle $\pi\pm\pi/N$ with each other. The resulting classical energy per bond is $E_{coplanar} /N  = \cos(\pi +\pi/N) JS^2$, yielding $-0.809017 JS^2 $ for the pentagon and $-0.900969 JS^2$ for the heptagon. These are in excellent agreement with the estimates obtained by fitting the numerical data. 

\begin{figure*}
\includegraphics[height = 10.5 cm]{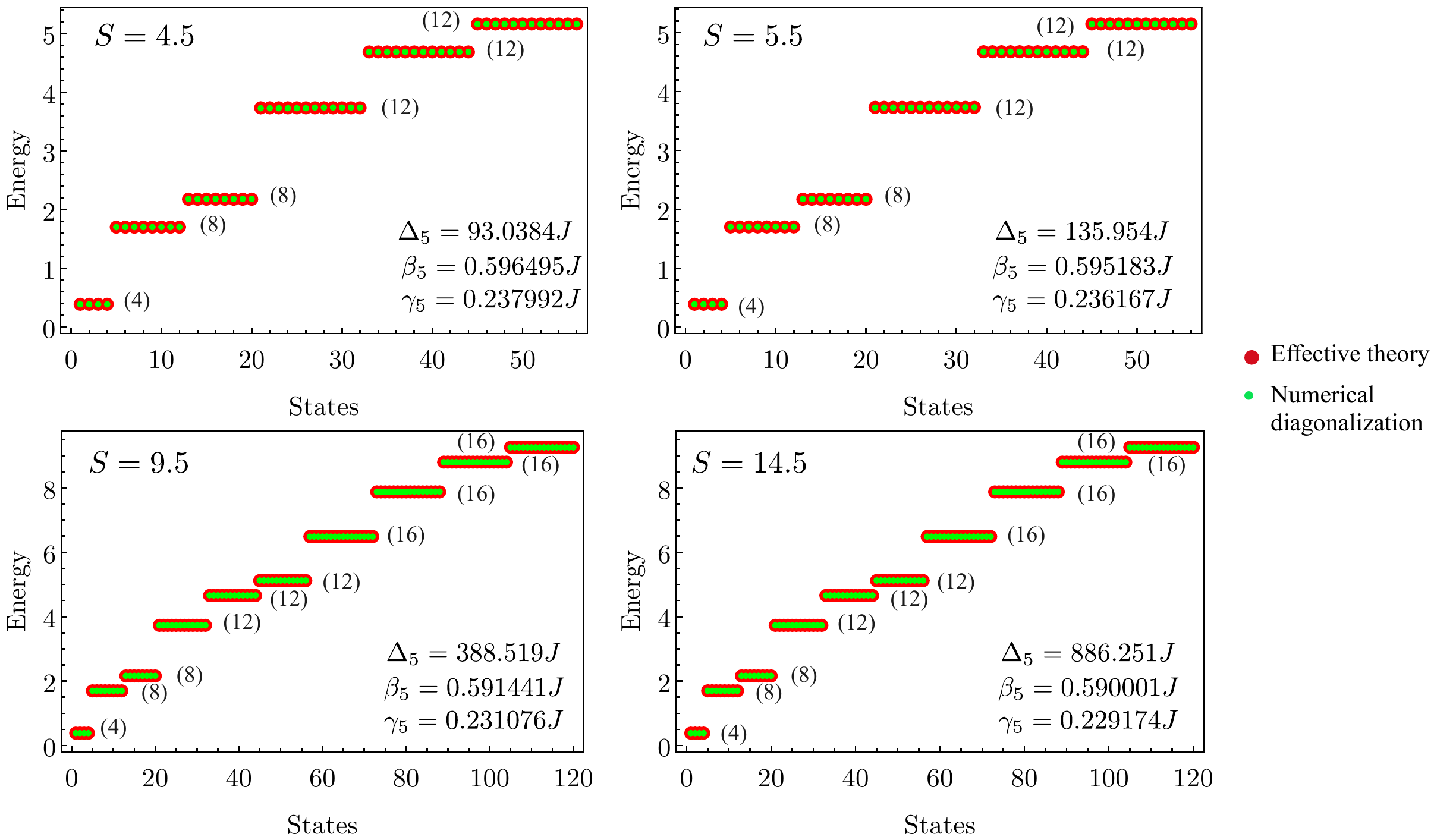}
\caption{Low-lying spectra of the pentagon antiferromagnet for various half-integer values of $S$. Plots show numerically-obtained spectra as well as fits to the effective theory. Fitting parameters are shown in each panel. Energies are measured in units of $J$. Numbers in parenthesis denote the degeneracy of each level. }
\label{fig.pent_half_int}
\end{figure*}

{\bf{Energy spectrum of pentagon:}} Fig.~\ref{fig.pent_half_int} shows numerically-obtained low-energy spectra for the pentagon $(N = 5)$ for various half-integer values of $S$. The low-lying levels show a degeneracy pattern that is consistent with the effective description -- that of a symmetric-top rigid rotor with a $\pi$-Berry phase. To perform a quantitative comparison, for each $S$, we fit the numerically-obtained spectrum (within a suitable low-energy window) to the form
\bea
E = -\Delta_5 + \beta_5 j(j+1) - \gamma_5 m^2,
\label{eq.form}
\eea
where $j=\frac{1}{2}, \frac{3}{2}, \frac{5}{2}$, etc. and $m=-j,-j+1,\ldots,j-1,j$. We attach a degeneracy of $(2j+1)$ to each $(j,m)$. The co-efficient $\Delta_5$ represents a constant shift. The rotor coefficients $\beta_5$ and $\gamma_5$ are found by a least-squares fit. As seen from Fig.~\ref{fig.pent_half_int}, we obtain excellent agreement with this form. Note that each panel only shows a low-energy window where the spectrum is in good agreement with Eq.~\ref{eq.form}. At higher energies, the numerically-obtained degeneracies deviate from the rigid-rotor pattern. 

\begin{figure}
\includegraphics[height=6 cm]{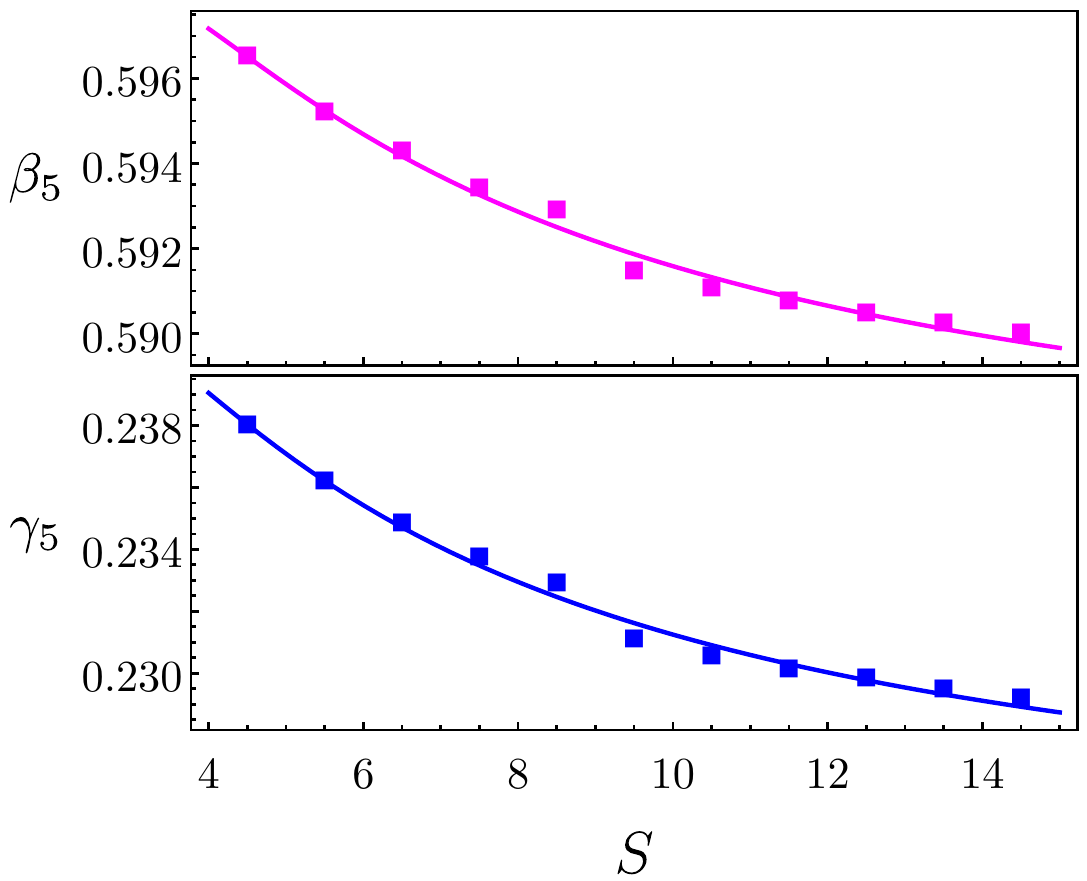}
\caption{$\beta_5$ and $\gamma_5$ vs. $S$. $\beta_5$ values are fit to the curve $\beta_5(S) = 0.58507 + 0.07628/S- 0.11165/S^2$. Those for $\gamma_5$ are fit to $\gamma_5(S) = 0.22286 + 0.09662/S-0.12751/S^2$. Both quantities are measured in units of $J$.}
\label{fig.pent_beta_gamma}
\end{figure}

The form in Eq.~\ref{eq.form} is motivated by the effective theory derived in Sec.~\ref{ssec.nlsm}. The analytically obtained values for the coefficients are $\beta_5^{eff.theory} = 0.58541J$ and $\gamma_5^{eff.theory} = 0.22361J$ (see App.~\ref{app.exps}). For all $S$, the values obtained by fitting the numerical data are close to the analytic result. As $S$ increases towards $\infty$, the fit values come closer to the analytic result. This can be seen in Fig.~\ref{fig.pent_beta_gamma}, where the $S$-dependence of $\beta_5$ and $\gamma_5$ is plotted. As described in the caption, they can be fit to polynomial forms to extrapolate to the $S\rightarrow\infty$ limit. This yields $\beta_5^{S\rightarrow \infty}=0.58507J$ and $\gamma_5^{S\rightarrow \infty}=0.22286J$ respectively. These values are very close to the analytic result, as we expect from the large-$S$ approach of the effective theory.

\begin{figure*}
\includegraphics[height = 5.3 cm]{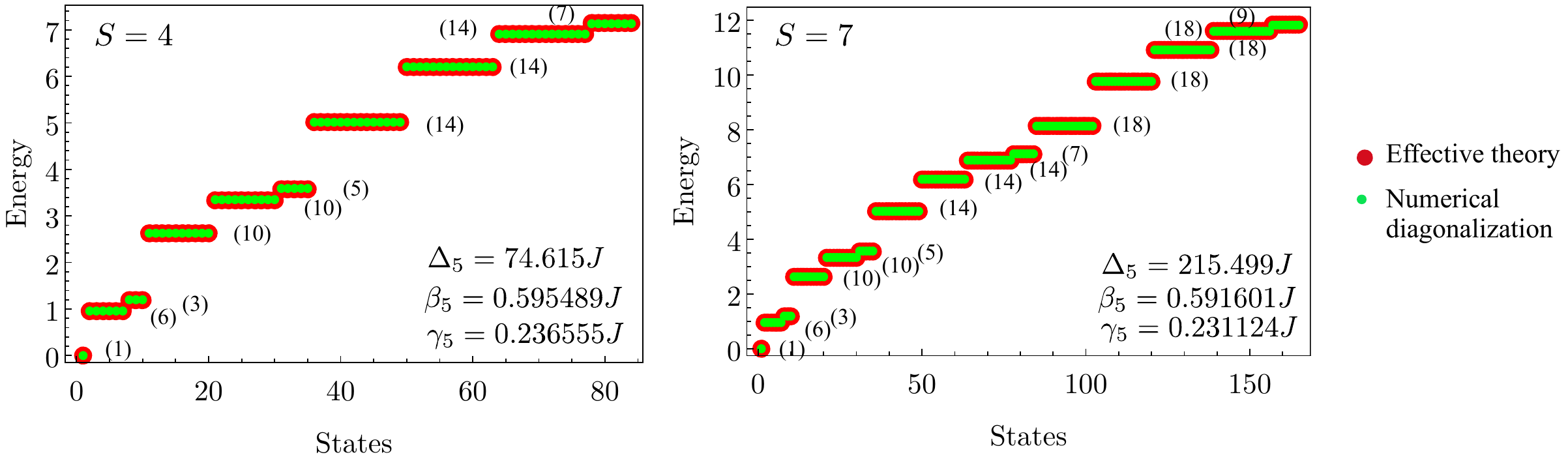}
\caption{Low energy spectra obtained from numerical diagonalization of the pentagon antiferromagnet for two integer values of $S$. Data have been fit to the rigid rotor form, with fitting parameters are shown in each panel. Energies are measured in units of $J$.}
\label{fig.pent_int}
\end{figure*}

\begin{figure*}
\includegraphics[height = 5.3 cm]{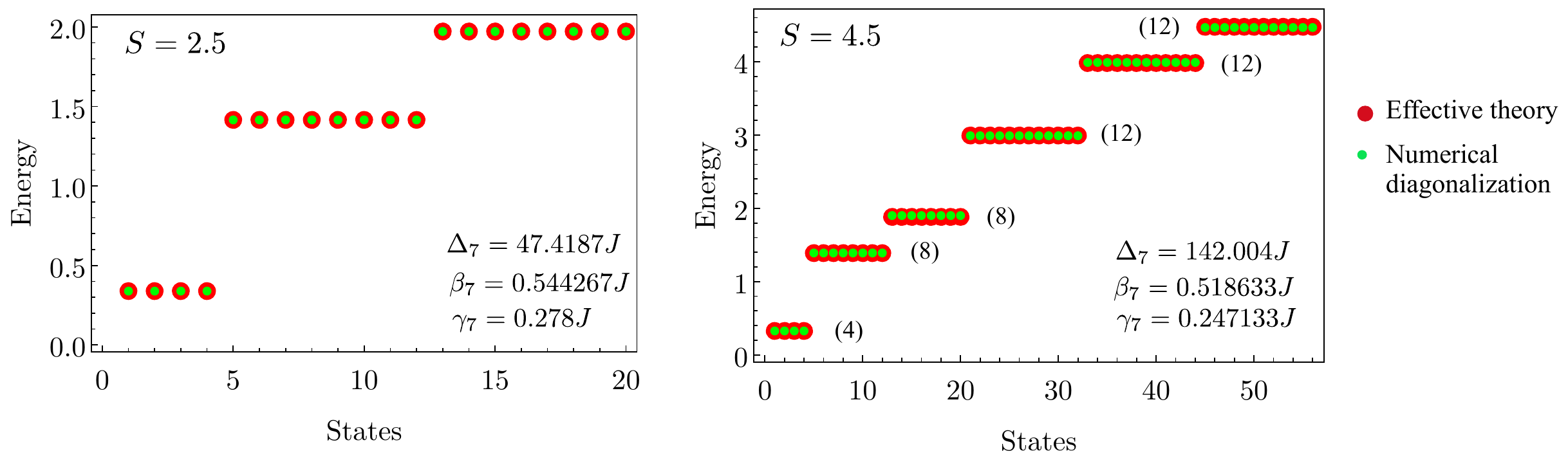}
\caption{Low energy spectra of the heptagon antiferromagnet for two half-integer values of $S$. Energies are measured in units of $J$.}
\label{fig.hept_half_int}
\end{figure*}

\begin{figure*}
\includegraphics[height = 5.3 cm]{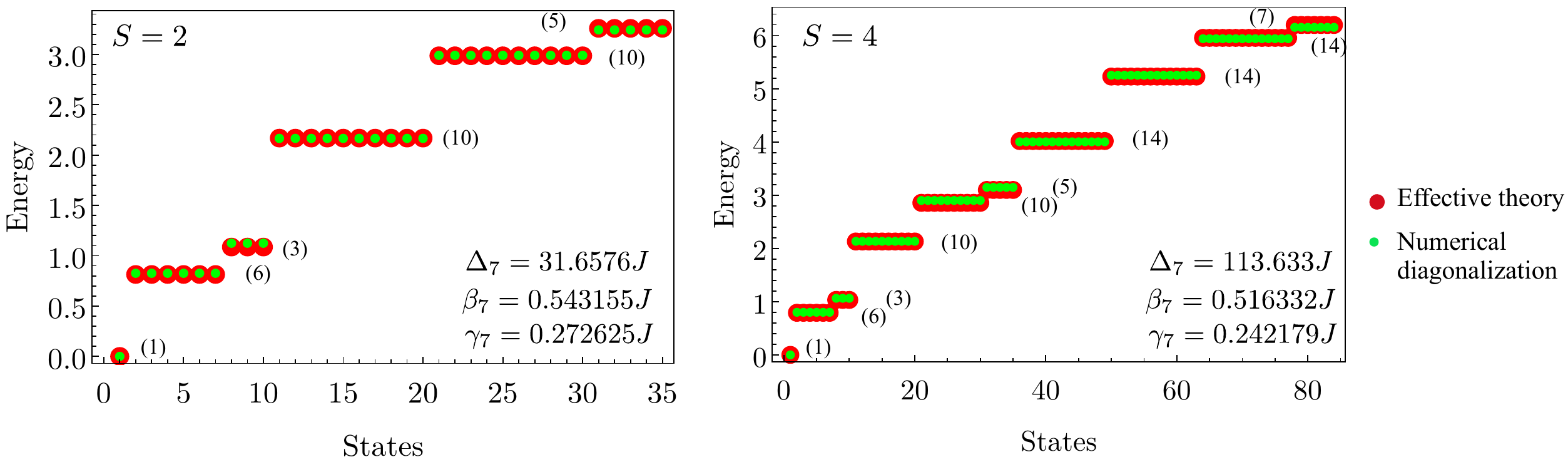}
\caption{Low energy spectra of the heptagon antiferromagnet for two integer values of $S$. Energies are measured in units of $J$. }
\label{fig.hept_int}
\end{figure*}

As a counterpoint, we present results for \textit{integer} values of $S$ in Fig.~\ref{fig.pent_int}. The spectra show excellent agreement with a symmetric-top rigid rotor \textit{without} Berry phases. Note that the degeneracy pattern is very different from those of half-integer $S$. The change in pattern is a direct manifestation of the Berry phase. 
\begin{figure*}
\includegraphics[height = 5.3 cm]{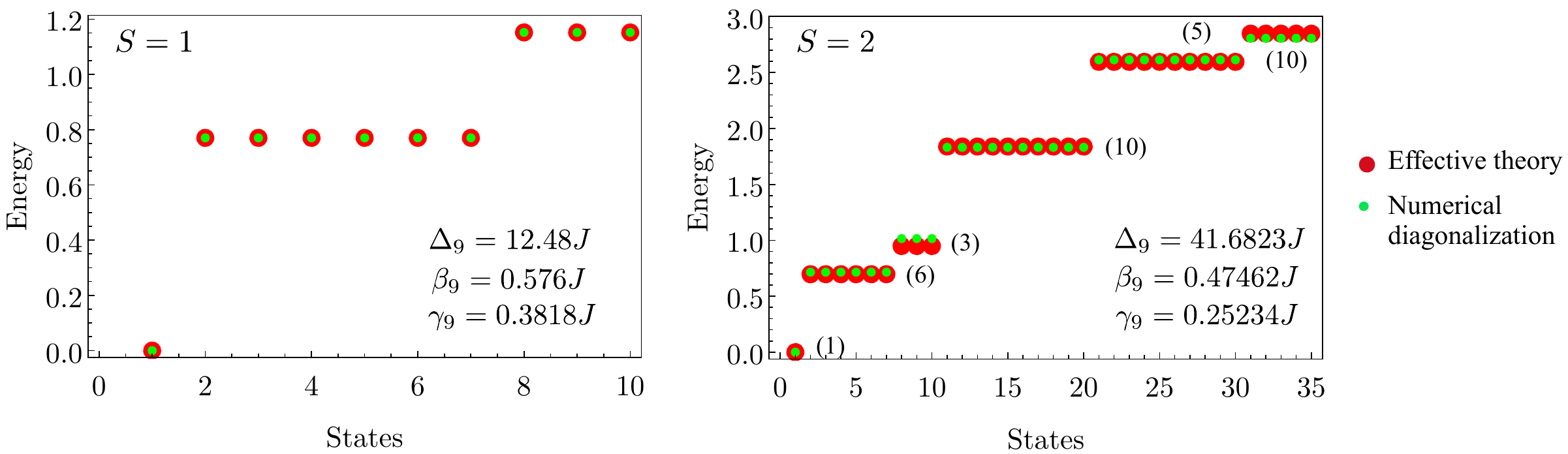}
\caption{Low energy spectra of the nonagon antiferromagnet for two integer spin values. Energies are measured in units of $J$.}
\label{fig.non_int}
\end{figure*}

\begin{figure}
\includegraphics[height = 5.3 cm]{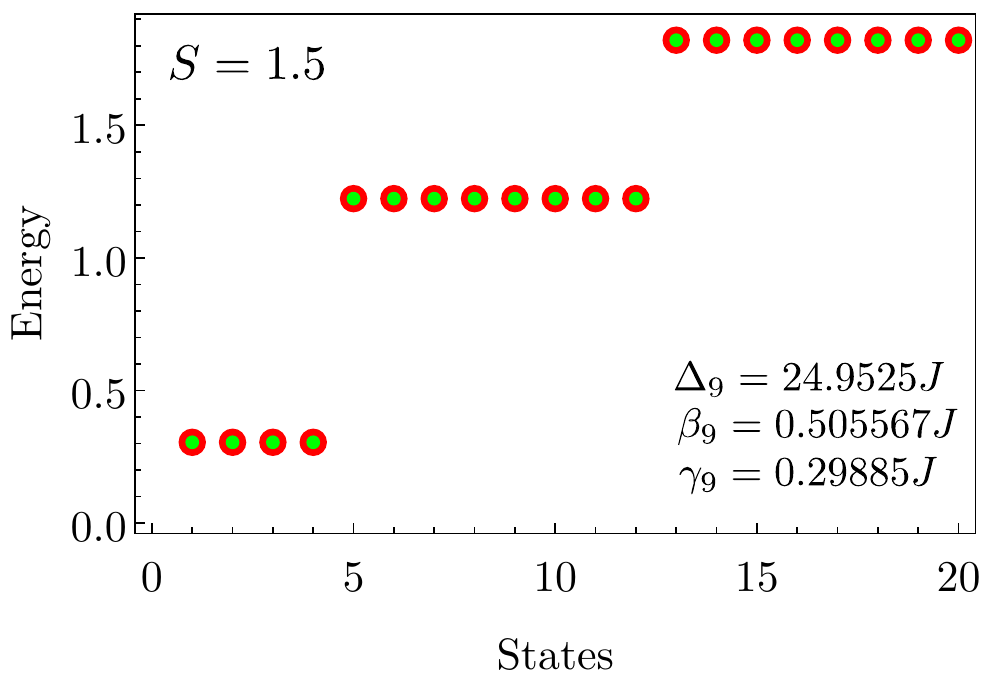}
\caption{Low energy spectrum of the nonagon antiferromagnet with $S = \frac{3}{2}$. Energies are measured in units of $J$.}
\label{fig.non_half_int}
\end{figure}

{\bf{Spectrum of larger polygons:}} We follow the same numerical approach and fitting procedure for heptagon ($N=7$) and nonagon ($N=9$) antiferromagnets. Heptagon spectra for half-integer and integer $S$ values are presented in Figs.~\ref{fig.hept_half_int} and \ref{fig.hept_int} respectively. Nonagon spectra with half-integer and integer spins are shown in Figs.~\ref{fig.non_int} and \ref{fig.non_half_int}. In all cases, spectra resemble those of a symmetric-top rotor. Integer-spins resemble the case with no Berry phase, while half-integer spins carry a Berry phase of $\pi$. For example, integer $S$ always leads to a non-degenerate ground state and a six-fold first excited state. In contrast, half-integer $S$ is invariably associated with a four-fold degenerate ground state and an eight-fold degenerate first excited state.  

\section{Discussion}
The topology of $SO(3)$ has drawn the attention of physicists for many decades. It is well known that the homotopy group is $\mathbb{Z}_2$, with two types of closed loops -- trivial and non-trivial. We may expect this rich topological structure to give rise to many observable consequences. However, relatively few examples are known. Perhaps the best known is the $\mathbb{Z}_2$ vortex. This represents a topological defect in $SO(3)$-field theories, with a textured $SO(3)$ order parameter in two dimensions\cite{Kawamura1984}. While this concept has been explored in many theoretical studies\cite{Kawamura2010,Rahmani2013,Osorio2019}, it has recently been invoked in experiments as well\cite{Tomiyasu2021}. Our results bring out a more direct consequence of $SO(3)$'s topology -- at the level of a single rotor, without requiring a field theory. Our results apply to a potentially large family of magnets. Our arguments regarding the Berry phase hold for any magnet that satisfies the three conditions listed in the abstract -- as emphasized in App.~\ref{app.Berry_phase}.

As specific examples, we have discussed odd-polygon antiferromagnets. Our discussion here builds upon previous studies on antiferromagnetic chains. Our results are consistent with early studies including analytic solutions for small spin values\cite{Kouzoudis1997} and observations from exact diagonalization spectra\cite{Schnack2000,Barwinkel2000a,Barwinkel2000b,Barwinkel2003}. Our analysis shows that features in low-energy spectra can be cleanly understood in terms of an effective rigid rotor description. These include degeneracy and momentum carried by the ground state.

Effective low-energy theories offer a starting point to understand spontaneous symmetry breaking\cite{Anderson1952}. Classical ordering requires breaking the symmetries of the system. However, this is not possible in any finite quantum system. Rather, we obtain a characteristic spectrum of states, called the `Anderson tower' or the `thin spectrum'. Our results can be viewed in this perspective, as determining the nature of the Anderson tower for non-collinear ordering in a class of magnets. The tower of states corresponds precisely to rigid rotor spectra. We provide an effective low-energy theory for polygons with $N$ vertices. We support this with an analytic calculation of the spectrum for $N=3$ and numerically obtained spectra for $N=5,7$ and $9$. 

A deeper view of our results reveals a propensity towards ordering when $N\rightarrow \infty$. In the effective rigid-rotor picture, we find that the moment of inertia increases with $N$. In the thermodynamic limit, we have a `massive' rotor that can be easily pinned by an external symmetry breaking field. Formally, this can be stated as a sequence of limits. If a symmetry breaking field is held at a fixed strength while $N$ is increased, the system will order. This order will persist if the external field is then smoothly taken to zero.  
Within this paradigm, our results show a qualitative difference between systems with integer and half-integer $S$. The nature of the low-lying spectra are different in the two cases. An interesting future direction is to examine whether this leads to observable differences in the approach to classical ordering.

The thin spectrum is of significant interest in numerical studies on quantum magnets\cite{Wietek2017}. It serves as a signature for classical ordering that is otherwise inaccessible in a finite system. This line of reasoning has played an important role in demonstrating the emergence of classical ordering in the Heisenberg antiferromagnet on the triangular lattice\cite{Bernu1992,Bernu1994}. Rigid-rotor based field theories have been used to determine the thin spectrum in triangle-based antiferromagnets, with the structure also appearing in the entanglement spectrum\cite{Rademaker2015}. 
Our study highlights the role of the Berry phase in this problem. This gives rise to an odd-even effect where the spectrum oscillates between two characteristic patterns as $S$ is varied.
\section*{Acknowledgement}
We thank G. Baskaran, R. Shankar, and Diptiman Sen for many insightful discussions.  

\appendix
\section{Berry phase in $SO(3)$ antiferromagnets}
\label{app.Berry_phase}
In the main text, we discuss a specific family of odd-polygon antiferromagnets. In this appendix, we evaluate the Berry phase for a more general case. We consider a generic quantum magnet for which the classical ground state space (CGSS) is isomorphic to $SO(3)$. We restrict our attention to magnets with unpolarized classical ground states, i.e., with no net magnetization. We will make use of these restrictions in the arguments below. We now determine the leading Berry phase contribution, the $\mathcal{O}(S)$ term in Eq.~\ref{eq.Berryexp} of the main text. 

At low energies, closed paths in configuration space can be viewed as smooth deformations of loops within the CGSS. As a result, every closed loop is topologically equivalent to a trajectory that lies entirely within $SO(3)$ space. This component within $SO(3)$ gives the dominant contribution to the Berry phase. As seen from Eq.~\ref{eq.Berryexp} of the main text, this component yields an $\mathcal{O}(S)$ contribution, while fluctuations out of $SO(3)$ contribute at $\mathcal{O}(S^0)$.

Below, we will first demonstrate that the Berry phase is a topological quantity -- robust to smooth deformations within $SO(3)$. We will argue that any pair of topologically-equivalent loops must yield the same Berry phase. We will next explicitly calculate the Berry phase for the two classes of loops within $SO(3)$. 

To show the topological nature of the Berry phase, we follow the approach given in Ref.~\onlinecite{Dombre1989}. The $\mathcal{O}(S)$ Berry phase accrued while traversing a loop is given by 
 \bea
 C = iS\int_0^\beta d\tau \sum_{j=1}^N\vec{A}(\hat{\Omega}_j)\cdot\partial_\tau(\hat{\Omega}_j)\vert_{\hat{\Omega}_j = R(\tau)\hat{n}_j},
 \label{eq.C}
 \eea
 where the index $j$ runs over each spin in the system, with the $j^\mathrm{th}$ spin oriented along the unit vector $\hat{\Omega}_j$. The orientation changes with time, given by $\hat{\Omega}_j=R(\tau)\hat{n}_j$, where $R(\tau)$ is a time-dependent $SO(3)$ rotation. The vector-valued function $\vec{A}$ satisfies $\epsilon_{\alpha\beta\gamma}\partial_\beta A^\gamma(\hat{\Omega}_j)= \Omega_{j\alpha}$. This can be viewed as the vector potential generated by a magnetic monopole at the origin.\cite{Auerbach}
 
Consider two proximate closed paths in $SO(3)$. We denote the first as $R_0(\tau)$ and the second as $R_0(\tau)+ \delta R_0(\tau)$, where we have included a small time-dependent deformation with respect to the first. We have
\begin{eqnarray}
C_{R_0(\tau)} &=& iS\int_0^\beta d\tau \sum_{j=1}^N\vec{A}(R_0\hat{n}_j)\cdot\partial_\tau(R_0\hat{n}_j),\nonumber\\
C_{R_0(\tau)+ \delta R_0(\tau)} &=& iS\int_0^\beta d\tau\sum_{j=1}^N \vec{A}((R_0 + \delta R_0)\hat{n}_j)\nonumber\\
&&\hspace{1.8cm}\cdot\partial_\tau((R_0 + \delta R_0)\hat{n}_j).
\end{eqnarray}
 Using $\epsilon_{\alpha\beta\gamma}\partial_\beta A^\gamma(\hat{\Omega}_j)\vert_{\hat{\Omega}_j = R\hat{n}_j} = (R\hat{n}_j)_\alpha$, we find the difference in $C$ to be
\begin{eqnarray}
\delta C &=& C_{R_0(\tau)+ \delta R_0(\tau)} - C_{R_0}(\tau)  \nonumber\\
&=& iS\int_0^\beta d\tau\; \epsilon_{\alpha\beta\gamma}R_0^{\gamma\delta} (\delta R_0)^{\alpha\nu}\partial_\tau R_0^{\beta\rho}\sum_{j=1}^N n_j^\delta n_j^\nu n_j^\rho\nonumber\\
& =& iS\int_0^\beta d\tau\; \epsilon_{\alpha\gamma\beta}(R_0^{-1}\partial_\tau R_0)^{\alpha\alpha'}(R_0^{-1}\delta R_0)^{\beta\beta'} T^{\alpha'\beta'\gamma},\nonumber\\
\end{eqnarray}
where $T^{\alpha'\beta'\gamma}\equiv\sum_{j=1}^N n_j^{\alpha'} n_j^{\beta'} n_j^\gamma $.
To arrive at the last step we have used the identity, $\epsilon_{\alpha\beta\gamma} R_{\alpha\alpha'}R_{\beta\beta'}R_{\gamma\gamma'} = \epsilon_{\alpha'\beta'\gamma'}$. Note that $T$ has the following properties. It is symmetric under the exchange of any two of the indices. In addition, it satisfies $\sum_\alpha T^{\alpha\alpha\beta} = 0$ as long as $\sum_j \hat{n}_j = 0$. If the magnet is unpolarized, we indeed have  $\sum_j \hat{n}_j = 0$.

Using these properties of $T$ and the anti-symmetric character of $(R_0^{-1}\partial_\tau R_0)$ and $(R_0^{-1}\delta R_0)$, we find that $\delta C$ vanishes. The $\mathcal{O}(S)$ Berry phase is the same for any two paths that are smooth deformations of each other! This immediately implies that the $\mathcal{O}(S)$ Berry phase is a topological quantity -- with the same value for all paths within each topological class.

As discussed in the main text, closed loops within $SO(3)$ fall under two classes -- trivial and non-trivial. By definition, trivial loops can be shrunk to a point. This corresponds to a rotation matrix that remains constant in time. As seen from Eq.~\ref{eq.C}, this yields $C=0$. By topological equivalence, $C$ is zero for every trivial loop. 

\begin{figure}
\includegraphics[width=3in]{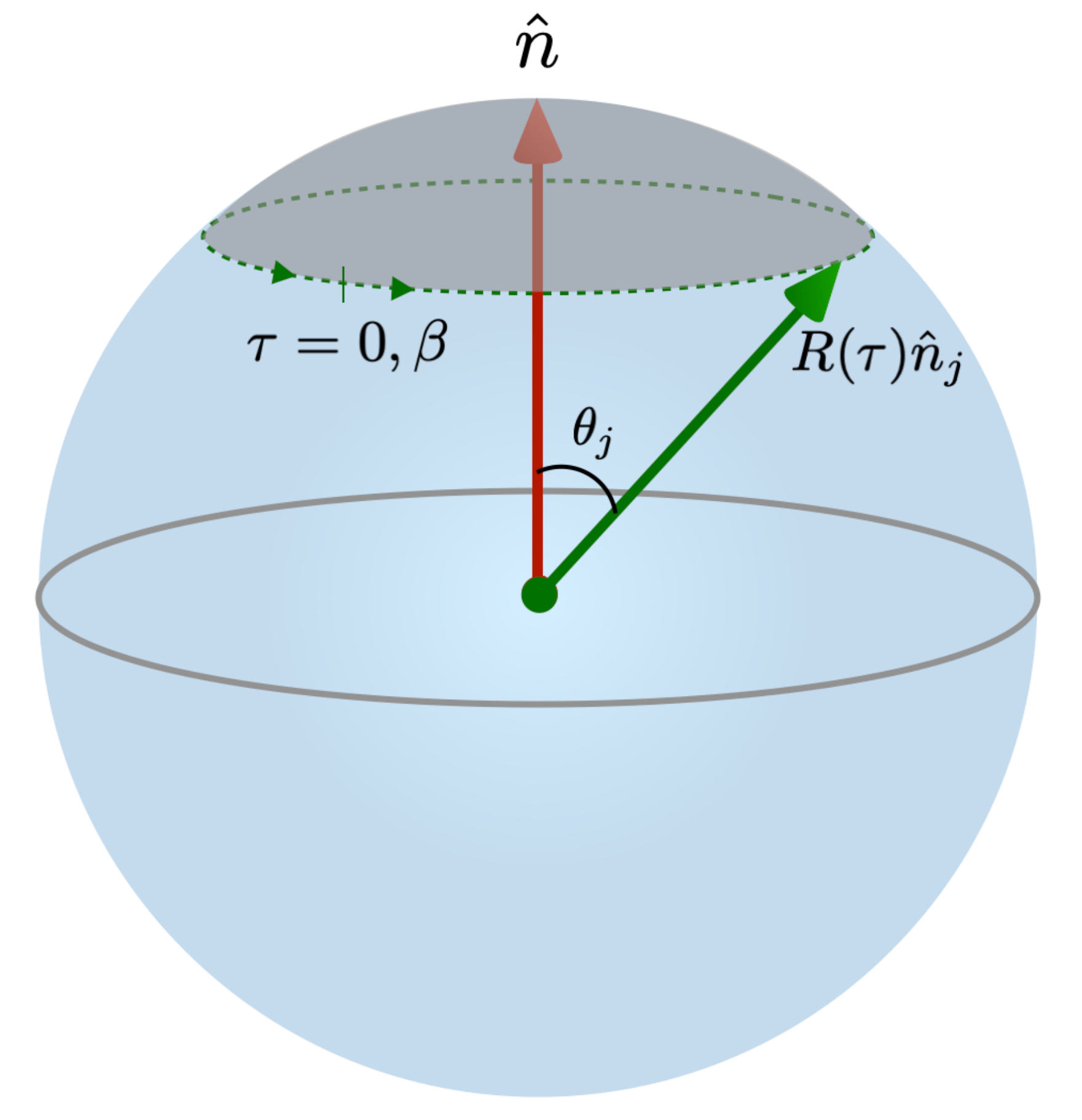}
\caption{The evolution of a spin as the system evolves along a non-trivial path. The axis of rotation is taken to be the north pole. As the spin rotates, it preserves the angle subtended with the axis, $\theta_j$. The subtended solid angle is shown in grey.}
\label{fig.theta}
\end{figure}

In the non-trivial class, we evaluate $C$ for a simple example -- a continuous family of rotations about an arbitrary axis, $\hat{n}$. At $\tau=0$, the angle of rotation is taken to be zero. At $\tau=\beta$, the angle is $2\pi$. At intermediate times, the angle increases linearly with time. This defines a closed loop in the non-trivial class for which the Berry phase can be easily evaluated. As we move along this path, each spin completes a rotation about the axis as shown in Fig.~\ref{fig.theta}. Note that the angle subtended with the axis, $\theta_j$, remains constant during time evolution. The Berry phase of Eq.~\ref{eq.C} can be recast in a geometric form\cite{Auerbach}
\bea
C = iS\sum_j \mathcal{A}_j,
\eea
where $\mathcal{A}_j$ is the solid angle subtended by the orientation vector of the $j^{th}$ spin. The solid angle may be defined with respect to an arbitrary point on the sphere. Here, we choose to measure solid angle from $\hat{n}$, giving $\mathcal{A}_j = 2\pi (1-\cos \theta_j)$. Adding the contribution from every spin, we obtain
\bea
C = i 2 \pi S \Big\{  N  -  \sum_j \cos \theta_j \Big\}.
\eea
To simplify this, we note that $ \cos\theta_j= \hat{n} \cdot \hat{n}_j$. We rewrite this as 
\bea
C = i 2 \pi S N  - i 2\pi  \hat{n} \cdot \Big\{ S \sum_j \hat{n}_j \Big\}.
\label{eq.Berryform1}
\eea
In the second term, the term within braces is the net magnetization of the system. In a magnet where the classical ground states are unpolarized, this term vanishes. We are left with $C = i 2\pi NS$ as the Berry phase. As non-trivial paths are topologically equivalent, they all accrue the same Berry phase.  

With these arguments, we obtain the leading $\mathcal{O}(S)$ contribution to the Berry phase given in Eq.~\ref{eq.Berryform} of the main text. 

\section{Periodicity of half-integer $j$ wavefunctions}
Wigner $D$ matrices are defined as
\bea
D_{m',m}^j (\hat{n},\theta) &=&  \langle j,m'\vert e^{-i\theta \hat{n}\cdot \vec{\hat{L}} } \vert j,m\rangle,
\eea
where $\vert j,m\rangle$'s are angular momentum eigenkets with $m$ and $m'$ being the z-projections of angular momentum. We now introduce $\vert j,j_n\rangle $'s, a new basis with momenta quantized along $\hat{n}$, rather than along $\hat{z}$. That is, they represent eigenstates of the operator $(\hat{n}\cdot \vec{L})$, with $(\hat{n}\cdot\vec{L}) \vert j, j_n\rangle = j_n \vert j,j_n\rangle $ and $j_n = -j,-(j-1),\cdots, (j-1), j$. 

We now re-express the $\vert j, m\rangle$'s as linear combinations of the new states. We write $\vert j,m\rangle = \sum_{j_n}c_{j_n}\vert j,j_n\rangle $. We obtain
 \bea
  e^{-i\theta \hat{n}\cdot \vec{\hat{L}} } \vert j,m\rangle = \sum_{j_n}c_n e^{-ij_n\theta}\vert j,j_n\rangle.
  \eea
 With this basis change, the Wigner $D$ matrices transform as
  \bea
  \langle j,m'\vert e^{-i\theta \hat{n}\cdot \vec{\hat{L}} } \vert j,m\rangle &=& \sum_{j_{n'}}\sum_{j_n}c^*_{n'}c_n e^{-ij_n\theta}\langle j,j_{n'}\vert j,j_n\rangle\nonumber\\
&=& \sum_{j_{n'}}\sum_{j_n}c^*_{n'}c_n e^{-ij_n\theta}\delta_{j_{n'}j_n} \nonumber\\
&=& \sum_{j_n} \vert c_n\vert ^2 e^{-ij_n\theta}.
  \eea
Here, $\theta$ is a continuous variable, representing the angle of rotation about the axis $\hat{n}$. Using the above expression, we may compare the difference in the Wigner $D$ matrix when $\theta$ changes by $2\pi$. We obtain
  \bea
  \nonumber D_{m',m}^j (\hat{n},\theta+2\pi) &=& \sum_{j_n} \vert c_n\vert ^2 e^{-ij_n\theta}e^{-i2\pi j_n} \\
  &=& \pm D_{m',m}^j (\hat{n},\theta) .
  \eea
We note that when $j$ is an integer, the $j_n$ quantum numbers take integer values so that the factor $e^{-i2\pi j_n}$ reduces to unity. This leads to the $+$ sign in the above equation. However, when $j$ is a half-integer,  $j_n$'s are also half-integers. As a result, $e^{-i2\pi j_n} = -1$. This leads to a negative sign above.

Representing the Wigner $D$ matrix as a wavefunction, $\psi(\hat{n},\theta)$, we obtain $\psi(\hat{n},\theta+2\pi) = (-1)^{2j} \psi(\hat{n},\theta)$. Taking $\theta = -\pi$, we write $\psi(\hat{n},-\pi) = (-1)^{2j} \psi(\hat{n},\pi)$. Note that a rotation corresponding to $(\hat{n},-\pi)$ can be written as one corresponding to $(-\hat{n},\pi)$. These arguments lead to Eq.~\ref{eq.jrelation} of the main text. 
  
\label{app.periodicity}

\section{Coefficients in the effective theory}
\label{app.exps}
We give explicit expressions for the quantities defined in Eq.~\ref{eq.efftheory} of the main text. These quantities define the effective theory for an odd-polygon antiferromagnet of size $N$. We have
\bea
\beta_N &=& \frac{4J\alpha_N}{N^2} - \frac{2J\cos \theta_N}{N},\\
 \gamma_N &=& \frac{4J\alpha_N}{N^2}- \frac{J}{N} - \frac{J\cos\theta_N}{N}.
 \eea
These expressions involve the quantity $\alpha_N$ given by
\bea
\nonumber \alpha_N = \frac{\cos(\theta_N)}{2}\Big[(N-1)\cos(\theta_N)+\cos\big((N-1)\theta_N\big)\Big].
\eea 
It can be seen straightaway that $\gamma_3 = 0$. For $N \rightarrow \infty$, we have $\alpha_N \rightarrow \frac{N}{2}$ and $\theta_N \rightarrow \pi$. This leads to $\beta_N \rightarrow \frac{4J}{N}$ and $\gamma_N\rightarrow \frac{2J}{N}$. Both $\beta_N$ and $\gamma_N$ fall off as $\sim\frac{1}{N}$ in the thermodynamic limit. However, their ratio approaches a finite value with $\frac{\gamma_N}{\beta_N}\rightarrow \frac{1}{2}$.

\bibliographystyle{apsrev4-1} 
\bibliography{oddgon.bib}
\end{document}